\documentclass[11pt]{amsart}
\numberwithin{equation}{section}
\numberwithin{figure}{section}
\numberwithin{table}{section}

\usepackage{macros}

\title{W-algebras of the Deligne-Cvitanovi\'{c} Exceptional series \\ and the minimal 3d $\CN=4$ SCFT}

\author{Thomas Creutzig${}^1$}
\address{${}^1$Department of Mathematics\\ FAU Erlangen\\ 91058 Erlangen, Germany}
\email{thomas.creutzig@fau.de}

\author{Niklas Garner${}^2$}
\address{${}^2$ Mathematical Institute\\ University of Oxford\\ Oxford, OX2 6GG, UK}
\email{niklas.garner@maths.ox.ac.uk}

\author{Byeonggi Go${}^3$}
\address{${}^3$Department of Physics\\ Korea Advanced Institute of Science and 
Technology\\ Daejeon 34141, Republic of Korea}
\email{wjdzm14@kaist.ac.kr}

\author{Heeyeon Kim${}^4$}
\address{${}^3$Department of Physics\\ Korea Advanced Institute of Science and Technology\\ Daejeon 34141, Republic of Korea}
\email{heeyeon.kim@kaist.ac.kr}

\begin{document}
	
\begin{abstract}
We propose a three-dimensional field theory construction that realizes the vertex algebras associated with the intermediate Lie algebras and the related $C_2$-cofinite minimal $W$-algebras of the Deligne-Cvitanović (DC) series as boundary algebras. The construction is based on the minimal three-dimensional $\CN=4$ superconformal field theory coupled to a topological field theory. For a Neumann-type boundary condition compatible with the topological $A$-twist, the algebra of boundary local operators realizes the minimal $W$-algebra $W_{-h^\vee/6}(\mathfrak{g},f_{\text{min}})$. While this boundary condition is not deformable to the $B$-twist, we argue that a holomorphic-topological ($HT^B$) twist instead realizes the level-one affine algebras of the intermediate Lie algebras, providing a uniform three-dimensional origin for these vertex algebra structures.
\end{abstract}

\maketitle
\tableofcontents

\section{Introduction}

The Deligne-Cvitanovi\'c (DC) exceptional series \cite{deligne1996serie}
\be
A_1 ~\subset~ A_2 ~\subset~ G_2 ~\subset~ D_4 ~\subset~ F_4 ~\subset~ E_6 ~\subset~ E_7 ~\subset~ E_8\ ,
\ee
is a distinguished sequence of simple Lie algebras that appears repeatedly across mathematics and physics, most notably in conformal field theory, where algebraic structures and modularity impose strong constraints. A concrete instance arises in the classification of two-character rational conformal field theories by Mukhi, Mathur and Sen (MMS) \cite{Mathur:1988na}, where solutions to modular linear differential equations (MLDE) are naturally realized by the characters of level-one affine algebras associated with Lie algebras in the DC series. 

Interestingly, the MMS classification has a "missing hole", an additional solution beyond the DC level-one affine algebras, which has been proposed to correspond to the character of the affine algebra $(E_{7\frac12})_1$. The Lie algebra $E_{7\frac12}$ is the intermediate Lie algebra between $E_7$ and $E_8$, introduced in \cite{landsberg2006sextonions} as a non-reductive Lie subalgebra of $E_8$. More generally, let $\mathfrak{g}$ be a simple Lie algebra equipped with the grading by the highest root,
\be
\mathfrak{g} = \mathfrak{g}_{-2} \oplus \mathfrak{g}_{-1} \oplus \mathfrak{g}_0 \oplus \mathfrak{g}_1 \oplus \mathfrak{g}_2\ .
\ee
The intermediate algebra of $\mathfrak{g}$ is then defined as
\be\label{intermediate lie algebra}
\mathfrak{g}_0' \oplus \mathfrak{g}_1 \oplus \mathfrak{g}_2\ ,
\ee
where $\mathfrak{g}_0'$ is the semi-simple subalgebra of $\mathfrak{g}_0$. 
The study of the corresponding vertex algebras was initiated in the work of Kawasetsu \cite{kawasetsu2014intermediate}, although much of their present structural understanding remains guided by character-level consistency checks \cite{Lee:2023owa, Lee:2024fxa}.

In this paper, we realize the vertex algebras associated with the intermediate Lie algebras of the DC series as boundary vertex algebras of a twisted three-dimensional $\CN=4$ superconformal field theory (SCFT). The theory that plays a central role is the so-called minimal $\CN=4$ SCFT, denoted $\CT_{\text{min}}$, first discussed in \cite{Gang:2018huc} as the simplest example of the rank-zero SCFT, i.e. one whose Coulomb and Higgs branches are both zero-dimensional. Boundary vertex algebras for the topological $A$- and $B$-twists of $\CT_{\text{min}}$ with various boundary conditions have been extensively studied recently \cite{Gang:2021hrd,Gang:2023rei,Ferrari:2023fez,Creutzig:2024ljv}, particularly in the context of the 4d SCFT/VOA correspondence \cite{Dedushenko:2023cvd,Gaiotto:2024ioj,Kim:2024dxu,ArabiArdehali:2024vli,ArabiArdehali:2024ysy}. As we shall see, the intermediate vertex algebras do not arise from a \emph{topological} twist, instead appearing in the holomorphic-topological (HT) or minimal twist studied in e.g. \cite{Costello:2020ndc}.

To realize the vertex algebras associated with the intermediate Lie algebras, 
we consider the minimal SCFT coupled to a topological field theory
\be
\overline{\CT}_{\text{min}} \times T_{\mathfrak g}\ ,
\ee
where $\overline\CT_{\text{min}}$ denotes the orientation reversal of ${\CT}_{\text{min}}$, and $T_{\mathfrak{g}}$ is the level-one pure Chern-Simons theory whose boundary supports the simple affine vertex operator algebra $L_1(\mathfrak{g})$. The two theories are coupled
via a mixed CS term between the R-symmetry of $\overline\CT_{\text{min}}$ and a $U(1)$ gauge symmetry of $T_{\mathfrak{g}}$. This coupling cancels the boundary gauge anomaly and ensures that the 
supersymmetric Neumann boundary condition for $\overline\CT_{\text{min}}$ is well defined. 

We propose that the coupled system admits a Neumann-type boundary condition deformable to the topological $A$-twist in the sense of \cite{Costello:2018fnz, Brunner:2021tfl}. Due to the nontrivial coupling between $T_\mathfrak{g}$ and $\overline{\CT}_{\text{min}}$, the conformal vector of $L_1(\mathfrak{g})$ in the twisted theory is modified from its standard form and becomes the so-called Urod-shifted conformal vector \cite{Bershtein:2013oka,Arakawa:2020oqo}. We argue that the resulting algebra of boundary local operators realizes the minimal $W$-algebra
\be
W_{-h^\vee/6}(\mathfrak{g}, f_{\text{min}})\ , 
\ee
where $h^\vee$ denotes the dual Coxeter number of $\mathfrak{g}$. For $\mathfrak{g}$ in the DC series, these coincide with the $C_2$-cofinite and rational $W$-algebras studied by Kawasetsu \cite{kawasetsu2018algebras}. We compute the half-index \cite{Gadde:2013sca, Yoshida:2014ssa, Dimofte:2017tpi} of the twisted theory, which counts BPS local operators on the boundary, and propose a family of novel Nahm sum formulas for their characters. In the case $\mathfrak{g}=\mathfrak{e}_8$, the characters of the twisted modules coincide with the modular invariant characters of $(E_{7\frac12})_1$ appearing in the MMS classification.  

On the other hand, we conjecture that the standard Neumann boundary condition does not admit a deformation compatible with the topological $B$-twist. One may instead consider the HT twist defined using $R_C$, the Cartan of $SU(2)_C$, i.e., part of the $R$-symmetry used to define the twisting homomorphism in the topological $B$-twist. We refer to this as the $HT^B$-twist, cf. \cite{Garner:2022vds}. In this setting, the algebra of boundary local operators can be computed explicitly, and we argue that it realizes the level-one affine algebras associated with the intermediate Lie algebras, including $(E_{7\frac12})_1$. The corresponding half-indices reproduce their conjectural character formulas discussed in \cite{kawasetsu2014intermediate,Lee:2024fxa}. 

The rest of the paper is organized as follows. In Section \ref{sec: Tmin}, we briefly review the basic properties of the minimal SCFT $\CT_{\text{min}}$ and its boundary conditions, leading to the construction of the coupled system $\overline\CT_{\text{min}}\times T_{\mathfrak{g}}$. In Section \ref{sec: sl2}, we analyze the simplest example with $\mathfrak{g}=\mathfrak{sl}_2$, where the boundary algebra realizes $W_{-1/3}(\mathfrak{sl}_2, f_{\text{min}})$, which in this case reduces to the Virasoro minimal model $M(5,3)$. In Section \ref{sec: T1 e8}, we examine the distinguished example $\mathfrak{g}=\mathfrak{e}_8$, for which the boundary degrees of freedom $L_1(\mathfrak{e}_8)$ form a holomorphic conformal field theory; this analysis leads to a novel level-rank duality between $M(5,2)$ and the minimal $W$ algebra $W_{-5}(\mathfrak{e}_8,f_{\text{min}})$. In Section \ref{sec: T2}, we consider a related construction involving the next simplest rank-zero theory $\CT_2$ coupled to $L_1(\mathfrak{e}_8)$, which is associated with an exotic intermediate algebra $(X_1)_1$. In Section \ref{sec: DC}, we treat the general case with $\mathfrak{g}$ in the DC series. Finally in Section \ref{sec: mirror}, we propose a mirror dual description $\mathbb{T}_\mathfrak{g}$, whose Dirichlet boundary conditions realize these boundary algebras, and discuss the deformability of the corresponding boundary conditions.

\subsection*{Acknowledgments} The work of N.G. was previously supported by ERC Consolidator Grant \#864828 “Algebraic Foundations of Supersymmetric Quantum Field Theory” (SCFTAlg) and is currently supported by Simons Collaboration on Celestial Holography CH-00001550-1. The work of B.G. and H.K. is supported by the National Research Foundation of Korea grant NRF2023R1A2C1004965 and RS-2024-00405629, and also by POSCO Science Fellowship of POSCO TJ Park Foundation.

\section{The minimal \texorpdfstring{$\CN=4$}{N=4} SCFT on half-space} \label{sec: Tmin}

In this section, we review some basic aspects of the minimal $\CN=4$ SCFT, $\CT_{\text{min}}$, which was first introduced in \cite{Gang:2018huc}.  Although the theory does not admit a Lagrangian description with manifest $\CN=4$ symmetry, it is conjectured to admit an ultraviolet $\CN=2$ realization in terms of a simple abelian Chern-Simons-matter theory. We then discuss its topological twists and various boundary conditions.

\subsection{\texorpdfstring{$\CN=2$}{N=2} Lagrangian description} Let us consider a $U(1)$ Chern-Simons theory with the level $k=3/2$, coupled to a chiral multiplet $\Phi$ of charge 1. It is argued in \cite{Gang:2018huc} that this theory flows in the infrared to the so-called minimal superconformal theory, $\CT_{\text{min}}$, with zero-dimensional Higgs and Coulomb branch. 

The UV gauge theory enjoys the global symmetry $U(1)_R \times U(1)_S$, where $U(1)_S$ can be identified with the topological symmetry associated with the $U(1)$ gauge group. In the infrared, the R-symmetry may mix with the topological symmetry; this mixing is parametrized by $\nu\in \mathbb{R}$, and we denote
\be
R_\nu = R_0 + \nu S\ ,
\ee
where we have chosen $R_0$ to be the superconformal R-charge at the fixed point, which is determined by F-maximization \cite{Jafferis:2010un}. This global symmetry is expected to enhance to the full R-symmetry group in the IR $\CN=4$ SCFT, $SO(4)\simeq SU(2)_C\times SU(2)_H/\mathbb{Z}_2$, with the embedding
\be
R_0 =  R_C + R_H \ ,\quad S = R_C - R_H\ ,
\ee
where $R_{C,H}$ are Cartan generators of $SU(2)_{C,H}$. To support this claim, one can perform a semiclassical analysis to argue that there exists two quarter-BPS, gauge invariant monopole operators that sit in the extra-supercurrent multiplets. They are
\be\label{quarter BPS monopoles}
\phi^2 V_{-1}\ ,\quad \bar\psi V_{+1}\ ,
\ee
whose superconformal $R$-charge and spin are both 1, with axial charge $S=-1$ and $1$, respectively. We denote the orientation reversal of $\CT_{\text{min}}$ by $\overline{\CT}_{\text{min}}$.

The $\CN=4$ SCFT $\CT_{\text{min}}$ can be topologically twisted to produce a pair of TFTs, which we denote by $\CT_{\text{min}}^A$ and $\CT_{\text{min}}^B$. For the $A$- and $B$-twist, the twisted spins are given by $J_A = J_3 + R_H$ and $J_B = J_3 + R_C$ respectively, corresponding to the choices $\nu=-1$ and $\nu=1$.  Although the boundary algebras of such non-Lagrangian TFTs are generally difficult to access, the existence of an $\CN=2$ Lagrangian description allows one to first pass to the holomorphic-topological (HT) twist and then study its deformation to a fully topological theory. This deformation is implemented by adding a proper superpotential term $\Theta_{A/B}$ to the twisted theory; the chiral operators appearing in this superpotential are precisely the local operators that are given in \eqref{quarter BPS monopoles}. See e.g. \cite{Ferrari:2023fez} for more details.

\subsection{Boundary conditions} \label{subsec: boundary conditions} The $\CN=2$ Lagrangian theory can be placed on a half-space $\mathbb{C}\times \mathbb{R}_{\leq 0}$ with a choice of half-BPS $(0,2)$ boundary condition. For such a boundary condition to be compatible with the topological $A$- or $B$-twist of the IR SCFT, it must be deformable to either $A$- or $B$-twist in the sense of \cite{Costello:2018fnz,Brunner:2021tfl}.

In \cite{Gang:2023rei}, it was argued that a deformation of a supersymmetric Dirichlet boundary condition is deformable to the $A$-twist, and the algebra of boundary local operators realizes the Virasoro minimal model $M(5,2)$. Similarly, an undeformed Dirichlet boundary condition was argued to be deformable to the $B$-twist in \cite{Ferrari:2023fez} and realizes the simple affine VOA $L_1(\mathfrak{osp}_{1|2})$. In a similar fashion, the orientation reversed theory $\overline{\CT}_\text{min}$ was argued in \cite{Ferrari:2023fez, Creutzig:2024ljv} to admit a supersymmetric Neumann boundary condition deformable to the $A$-twist and an appropriate (though its explicit form is unknown) deformation thereof deformable to the $B$-twist.

The precise Neumann boundary condition is constrained by the absence of gauge anomalies. Let ${\bf f}$, $\bf r$ and ${\bf f}_{\text{top}}$ be the field strength for the gauge, $R$, and topological symmetries, respectively. The perturbative boundary anomaly of the (orientation reversed) UV gauge theory is then \cite{Dimofte:2017tpi,Ferrari:2023fez}
\be\label{gauge anomaly}
-2 {\bf f}^2 - 2({\bf f}_{\text{top}}-{\bf r}){\bf f}\ ,
\ee
which must be canceled by adding appropriate (0,2) degrees of freedom $T_{2d}$ at the boundary. Notice that the mixed anomaly between the gauge-global symmetry above must cancel as well for the deformation to the $A$-twist, otherwise the global symmetry $R_H$, which we need for the topological twisting, is broken. 

A natural choice of $T_{2d}$ is the algebra of two complex Majorana fermions, each comprising a (0,2) Fermi multiplet. As discussed in \cite{Ferrari:2023fez,Creutzig:2024ljv}, one may assign gauge charge 1 to one Fermi multiplet, and gauge charge $-1$, topological charge $-1$, and R-charge $1$ to the second Fermi multiplet, to completely cancel the gauge anomalies \eqref{gauge anomaly}.

The half-index counting boundary local operators is defined by
\be
I_\text{half}(q;x) = \text{tr}_{\text{ops}} (-1)^{R_\nu} q^{J_3 + R_\nu/2} x^F\ ,
\ee
where $F$ is generator of the boundary flavor symmetry and $q=e^{2\pi i\tau}$. The Neumann half-index of $\overline{\CT}_{\text{min}}$ with the above Fermi multiplets reads
\be
I_\text{half}(q;x,\nu,\eta) = (q)_\infty\oint \frac{dz}{2\pi i z} \frac{FF(z x)FF((-q^{1/2})^{\nu-1}\eta q zx^{-1})}{(z,q)_\infty} \ ,
\ee
where $FF(z) = (z;q)_\infty (qz^{-1};q)_\infty$ is the elliptic genus of a boundary $\CN=(0,2)$ Fermi multiplet. Here $x$ is the fugacity for the $U(1)$ global symmetry under which the two fermions have charge 1, and $\eta$ counts the charges under the global symmetry $S$. The $A$-twist specialization ($\nu=-1$, $\eta=1$) reproduces the vacuum character of $L_1(\mathfrak{osp}_{1|2})$, while the $B$-twist specialization ($\nu=1$, $\eta=1$) together with $x=1$ reproduces the vacuum character of $M(5,2)$.

The perturbative anomalies of 2d $\CN=(0,2)$ boundary degrees of freedom are encoded in the transformation of the elliptic genus under large gauge transformations \cite{Alvarez-Gaume:1986rcs} (see also \cite{Closset:2019ucb}). Suppose that the 2d degrees of freedom have $G=U(1)^N$ global symmetry with corresponding fugacities $x_a=e^{2\pi i u_a}$, $a=1,\cdots, N$. Then under a large gauge transformation, $u\rightarrow u+m\tau$ for $m\in \mathbb{Z}^N$, the elliptic genus transforms as
\be\label{large gauge}
I_{2d}(q;\{x_a\}) \rightarrow e^{-\pi i \CA^{ab}(m_a u_b + u_a m_b + m_am_b\tau) }I_{2d}(q,\{x_a\})\ ,
\ee
where $\CA^{ab}{\bf f}_a{\bf f}_b$ is the anomaly polynomial of the $G$-symmetry. Indeed one can check that the contribution from the boundary fermions,
\be
I_{2d}(q;z) = FF(z)FF((-q^{1/2})^{\nu-1}\eta q z)
\ee
transforms as
\be
I_{2d}(q;zq) = q^{-1}(-q^{1/2})^{1-\nu} \eta^{-1}z^{-2} I_{2d}(q;z)\ ,
\ee
which is consistent with \eqref{large gauge}.

Alternatively, the gauge anomalies \eqref{gauge anomaly} can also be canceled by taking the boundary degrees of freedom to be a level-one chiral WZW model, giving rise to the simple affine vertex algebra $L_1({\mathfrak g})$. In general, this does not define a 2d holomorphic CFT, i.e. they are not genuinely 2d degrees of freedom.\footnote{An interesting exception occurs for $\mathfrak{g}=\mathfrak{e}_8$, which we discuss separately in Section \ref{sec: T1 e8}} Instead, they are most naturally interpreted as edge modes for a 3d Chern-Simons theory with gauge group $G$ at level 1, taken to be the simply connected group with Lie algebra $\mathfrak{g}$, together with Dirichlet boundary conditions. 
This leads us to consider the boundary algebra of the coupled system $\overline{\CT}_{\text{min}}^A\times T_{\mathfrak{g}}$ or, equivalently, an interface between $\CT^A_{\text{min}}$ and $T_{\mathfrak{g}}$. 

Let us consider the $\mathfrak u(1)$ subalgebra of $\mathfrak{g}$ generated by $h_\theta = \theta^\vee$, which induces the grading
\be
\mathfrak{g} = {\mathfrak g}_{-2}\oplus {\mathfrak g}_{-1} \oplus {\mathfrak g}_{0} \oplus {\mathfrak g}_{1} \oplus {\mathfrak g}_{2}\ ,
\ee
and let $\mathfrak{g}_0'$ be the semisimple part of $\mathfrak{g}_0$.
The two bulk theories $\overline{\CT}^A_{\text{min}}$ and $T_{\mathfrak{g}}$ are coupled at the boundary by identifying the subalgebra generated by $h_\theta$ with the boundary $\mathfrak{u}(1)$ gauge symmetry of $\overline{\CT}^A_{\text{min}}$. Then the corresponding current $J_{h_\theta}$ has level 2, 
\be
J_{h_\theta}(z) J_{h_\theta}(w) \sim \frac{2}{(z-w)^2}\ ,
\ee
providing exactly the contribution needed to cancel the pure gauge anomaly from $\overline{\CT}_{\text{min}}$.

In order to generate the desired mixed gauge-global anomalies, we introduce a coupling between the bulk QFTs in the form of a mixed CS term
\be\label{mixed gauge global}
\frac{1}{2\pi}\int (A_{\text{top}} - A_{R})\wedge dA\ ,
\ee
where $A_{\text{top}}$ and $A_R$ are background gauge fields for the topological and $R$-symmetries of $\CT_{\text{min}}$ respectively, while $A$ is the $h_\theta$-gauge field in $T_{\mathfrak{g}}$. At the interface, the latter is identified with that of the $U(1)$ gauge field on the $\CT_{\text{min}}$ side.

After taking the topological $A$-twist, the presence of the mixed CS term \eqref{mixed gauge global} induces an effective shift of the topological spins of the Wilson lines in $T_\mathfrak{g}$, proportional to their charges under $h_\theta$. To see this, consider a Wilson line of charge $n$ in the level-2 $U(1)$ CS theory associated with the $h_\theta$ gauge field. The expectation value of the Wilson line can be computed by the path integral
\be
\langle W_n\rangle = \int [dA]_{\CA/\CG}~e^{\frac{i}{2\pi} \int A dA + \frac{i}{2\pi} \int w_h dA + in \oint_\gamma A}\ ,
\ee
where $w_h$ is the Cartan component of the spin connection $w$, associated with the rotation in the holomorphic plane in the $HT$-twist of the UV theory. The second term in the action originates from the mixed CS term \eqref{mixed gauge global} after the twist, under which the spin connection $w$ is identified with the connection $w_{R_H}$ on the principal $SU(2)$ bundle for the $R_H$-symmetry in the infrared. The classical value for the gauge field solves the equation of motion
\be
dA = -\frac12 dw_h -n\pi \delta_{\gamma}\ , 
\ee
where $\delta_\gamma$ is a Poincar\'e dual of the curve $\gamma$. Plugging it back, we find that the path integral reduces to
\be
\langle W_n \rangle = \exp\left[-\frac{i\pi n^2}{2} \int_D \delta_\gamma - \frac{in}{2} \int_D dw_h\right] \cdot Z\ ,
\ee
with $\partial D = \gamma$, where $Z$ is a $n$-independent factor. The first factor in the exponent is proportional to the self-linking number of the line, which is responsible for the framing anomaly. If we change the framing of the line by an integer $N$, the expectation value of the Wilson loop picks up a factor
\be
\langle W_N\rangle \rightarrow e^{i2\pi N\left(\frac{n^2}{4} + \frac{n}{2}\right) }\langle W_N\rangle\ ,
\ee
which implies that the topological spin of $W_n$ is
\be
\theta_n = \exp 2\pi i\left(\frac{n^2}{4} + \frac{n}{2}\right)\ .
\ee
The additional shift of $n/2$ in the exponent is due to the coupling \eqref{mixed gauge global}.

At the level of the boundary algebra, the coupling \eqref{mixed gauge global} effectively assigns the topological and R-charge of the boundary monopole operators in $T_{\mathfrak{g}}$ to be equal to $n$ and $-n$ respectively, where $n\in \mathbb{Z}$ is their magnetic flux under the $U(1)$ gauge symmetry. In the $A$-twist, this induces a shift of the twisted spins of the corresponding modules by $n/2$, which is essential for identifying the boundary edge modes for the Chern-Simons fields as an Urod-shifted simple affine Lie algebra, as we discuss in more detail in Section \ref{sec: sl2}.

In the following sections, we discuss a class of vertex algebras obtained in this way, when ${\mathfrak g}$ belongs to Deligne–Cvitanovi\'{c} exceptional series, and argue that this construction naturally leads to a physical realization of a class of Urod shifted W-algebras associated to $\mathfrak g$.

\section{The Urod phenomenon for \texorpdfstring{$\mathfrak{sl}_2$}{sl2}}\label{sec: sl2}

As a warm-up, we begin with the simplest case, $\mathfrak{g}=\mathfrak{sl}_2$. We study the boundary algebra of the coupled system $\overline{\CT}_{\text{min}}^A\times T_{\mathfrak{sl}_2}$, where $T_{\mathfrak{sl}_2}$ is the pure $SU(2)$ Chern-Simons theory at level 1. By imposing a Neumann boundary condition for $\overline{\CT}_{\text{min}}^A$ and a Dirichlet boundary condition for  $T_{\mathfrak{sl}_2}$, coupled in an appropriate way,
we find strong evidence that the resulting algebra of boundary local operators corresponds to the Virasoro minimal model $M(5,3)$. This construction may be viewed as an instance of the Urod phenomenon for $L_1(\mathfrak{sl}_2)$\cite{Bershtein:2013oka}, which we review briefly in Section \ref{sec:urod sl2}.

\subsection{Coupling \texorpdfstring{$L_1(\mathfrak{sl}_2)$}{L1sl2} to \texorpdfstring{$\overline\CT_{\text{min}}^A$}{TminA}} \label{sec: coupling sl2}
Let $A$ denote the abelian gauge fields in $T_{\mathfrak{sl}_2}$, obtained by projecting the $\mathfrak{su}(2)$ connection onto its Cartan subalgebra. We couple this theory to $\overline{\CT}_{\text{min}}$ at the boundary by identifying the boundary value of $A$ with the dynamical gauge field of $\overline{\CT}_{\text{min}}$. Taking into account the mixed CS term \eqref{mixed gauge global}, the contribution from $T_{\mathfrak{sl}_2}$ is given by its Dirichlet half-index \cite{Dimofte:2017tpi},
\be
I[T_{\mathfrak{sl}_2}](q;z,\nu,\eta) = \frac{1}{(q)_\infty} \sum_{n\in \mathbb{Z}} q^{n^2}\left[(-q^{1/2})^{\nu-1}\eta\right]^{-n}z^{-2n}\ ,
\ee
where $z$ is the fugacity/Jacobi variable for the Cartan. One can check that this expression has the desired transformation properties with respect to large gauge transformations:
\be
I[T_{\mathfrak{sl}_2}](q;qz,\nu,\eta) =q^{-1}(-q^{1/2})^{1-\nu} \eta^{-1}z^{-2}  I_{2d}(q;z,\nu,\eta) \ .
\ee
It therefore completely cancels the anomaly inflow \eqref{gauge anomaly} from $\overline{\CT}_{\text{min}}$. 
In the $A$-twist specialization, the index simplifies to
\be
I[T_{\mathfrak{sl}_2}](q;z,-1,1) = \frac{1}{(q)_\infty} \sum_{n\in \mathbb{Z}} q^{n^2+n}z^{-2n}\ .
\ee
Calculating the half-index of the coupled system, we find 
\begin{equation}
    \begin{split}
    I_\text{half}(q,-1,1) & =(q)_\infty\oint \frac{dz}{2\pi i z} \frac{I[T_{\mathfrak{sl}_2}](q;z,-1,1) }{(z;q)_\infty} \\
 &   = \oint \frac{dz}{2\pi i z} \frac{1}{(z;q)_\infty}\sum_{n\in \mathbb{Z}} q^{n^2+n}z^{-2n} \\
& = \sum_{n\in \mathbb{Z}} \frac{q^{n^2+n}}{(q)_{2n}}\ .
\end{split}
\end{equation}
This coincides with the vacuum character of $M(5,3)$, which supports the proposal that the corresponding boundary vertex operator algebra is $M(5,3)$.

The bulk TFT admits simple lines $W_{(a,b)}$ with $a,b=0,1$, which can be identified as Wilson lines of charge $(a,b)$ under $U(1)_{\overline{\CT}_\text{min}}\times U(1)_{T_\mathfrak{g}}$ gauge symmetry in the UV description. We find that insertion of these lines reproduces the character of the four simple modules of $M(5,3)$,
\begin{equation}
\begin{split}
I_\text{half}(q,-1,1)[W_{(1,0)}] &= \oint \frac{dz}{2\pi i z} \frac{1}{(z;q)_\infty}\sum_{n\in \mathbb{Z}} q^{n^2+n}z^{-2n} z^{-1}\\
& = \sum_{n\in \mathbb{Z}} \frac{q^{n^2+n}}{(q)_{2n+1}} = q^{-9/40}\text{ch}(M_{3,1}^{5,3})\ ,
\end{split}
\end{equation}
\begin{equation}
\begin{split}
I_\text{half}(q,-1,1)[W_{(0,1)}] &= \oint \frac{dz}{2\pi i z} \frac{1}{(z;q)_\infty}\sum_{n\in \mathbb{Z}} q^{n^2+2n+1}z^{-2n} z^{-1}\\
& = \sum_{n\in \mathbb{Z}} \frac{q^{(n+1)^2}}{(q)_{2n+1}} = q^{9/40}\text{ch}(M_{4,1}^{5,3})\ ,
\end{split}
\end{equation}
and
\begin{equation}
\begin{split}
I_\text{half}(q,-1,1)[W_{(1,1)}] &= \oint \frac{dz}{2\pi i z} \frac{1}{(z;q)_\infty}\sum_{n\in \mathbb{Z}} q^{n^2+2n+1}z^{-2n} z^{-2}\\
& = \sum_{n\in \mathbb{Z}} \frac{q^{n^2}}{(q)_{2n}} = q^{1/40}\text{ch}(M_{2,1}^{5,3})\ . 
\end{split}
\end{equation}

\subsection{Decomposition of \texorpdfstring{$L_1(\mathfrak{sl}_2)$}{L1sl2}} \label{sec:urod sl2}
The appearance of $M(5,3)$ can be understood from the Urod-phenomenon for $\mathfrak{sl}_2$, which says that the simple affine vertex algebra of $\mathfrak{sl}_2$ at level one is a conformal extension of two Virasoro minimal models. \cite{Bershtein:2013oka,Arakawa:2020oqo}

For this let $L_1(\mathfrak{sl_2})$ be the simple affine vertex algebra of $\mathfrak{sl}_2$ at level one and let $L_1(\omega)$ be the simple module whose top level is the standard representation of $\mathfrak{sl}_2$. 
Let $M(u, v)$ be the simple Virasoro algebra at central charge 
\be
c = 1 - 6(u-v)^2/uv\ . 
\ee
It is strongly rational for $u,v \in \mathbb Z_{>1}$ coprime. 
Its simple modules are $M^{u, v}_{r, s}$ for $1 \leq r \leq u-1, 1 \leq s \leq v-1$ and $M^{u, v}_{r, s} \cong M^{u,v}_{u-r, v-s}$. Then the Urod phenomenon says, that
\begin{equation} \label{decomposition sl2}
    \begin{split}
  L_1(\mathfrak{sl_2}) &\cong M(5, 2) \otimes M(5, 3) \oplus  M^{5, 2}_{3, 1} \otimes    M^{5, 3}_{3, 1}\ , \\ 
  L_1(\omega) &\cong  M^{5, 2}_{2, 1} \otimes    M^{5, 3}_{2, 1} \oplus  M^{5, 2}_{4, 1} \otimes    M^{5, 3}_{4, 1}\ .
    \end{split}
\end{equation}
The stress tensor/conformal vector in this identification is \emph{not} the standard one coming from the Sugawara construction, witnessed by the fact that the central charge of the right-hand side is $-5$. Instead, it is a different stress tensor coming from the Urod conformal vector. We denote characters with respect to the conformal weight grading for the Urod conformal vector by a superscript $U$. The relation to usual characters is
\[
\text{ch}^U[L_1(\mathfrak{sl}_2)](q;z) = \text{ch}[L_1(\omega)](q;z), \qquad 
\text{ch}^U[L_1(\omega)](q;z) = \text{ch}[L_1(\mathfrak{sl}_2)](q;z)\ .
\]
Indeed we find that the contribution of $T_{\mathfrak{sl}_2}$ in the $A$-twisted limit agrees with the Urod-shifted character, up to a modular anomaly pre-factor:
\[
I[T_{\mathfrak{sl}_2}](q;z,-1,1) = \text{ch}^U[L_1(\mathfrak{sl}_2)](q;z)\ . 
\]

The decomposition \eqref{decomposition sl2} admits a natural interpretation in terms of the vertex algebra living at the interface between $\CT_{\text{min}}^A$ and $T_{\mathfrak{sl}_2}$. Consider $\CT_{\text{min}}^A$ on an interval $x_3\in[0,L]$, with a Dirichlet boundary condition at the right endpoint $x_3=L$ realizing $M(5,2)$ as in \cite{Gang:2023rei}, and a Neumann boundary condition at the left endpoint $x_3=0$. The left boundary is then glued to $T_{\mathfrak{sl}_2}$ defined on the half-space $x_3\leq 0$ with Dirichlet boundary condition, in the manner described in Section \ref{sec: coupling sl2}.  The interval reduction of $\CT_{\text{min}}^A$ identifies the extension of $M(5,2)\otimes M(5,3)$ with the boundary algebra of  $T_{\mathfrak{sl}_2}$ at $x_3=0$, which is precisely the Urod-shifted $L_1(\mathfrak{sl}_2)$.


\subsection{\texorpdfstring{$HT^B$}{HTB}-twist} \label{sec: HTB sl2} As discussed in \cite{Ferrari:2023fez,Creutzig:2024ljv}, the standard Neumann boundary condition for $\overline{\CT}_{\text{min}}$ is not deformable to the $B$-twist, since the superpotential term $\phi^2 V_{-1}$ implementing the deformation does not vanish at the boundary. Instead, one may consider the $HT^B$-twisted theory, namely the holomorphic-topologically twisted theory with respect to the $\CN=2$ $R$-charge given by $R= R_C$, cf. \cite{Garner:2022vds}. The Neumann half-index coupled to $L_1(\mathfrak{sl}_2)$ in the limit $\nu=\eta=1$ becomes
\begin{equation}
    \begin{split}
    I_\text{half}(q,1,1) & =(q)_\infty\oint \frac{dz}{2\pi i z} \frac{I[T_{\mathfrak{sl}_2}](q;z,1,1) }{(z;q)_\infty} \\
 &   = \oint \frac{dz}{2\pi i z} \frac{1}{(z;q)_\infty}\sum_{n\in \mathbb{Z}} q^{n^2}z^{-2n} \\
& = \sum_{n\in \mathbb{Z}} \frac{q^{n^2}}{(q)_{2n}}\ ,
\end{split}
\end{equation}
where the contribution from $L_1(\mathfrak{sl}_2)$ is now given by that with the standard conformal vector. In the absence of the superpotential term implementing the deformation to the $B$-twist, one can explicitly calculate the boundary operator algebra. Let $J_\pm, J_3$ be the generators for $L_1(\mathfrak{sl}_2)$. The boundary algebra is generated by the gauge invariant combinations of $\phi(z)$, $J_a(z)$, and derivatives of $c(z)$, together with the BRST differential $Q$ which acts as
\be
Q = \oint \frac{dz}{2\pi i} ~c(z) J_3(z)\ .
\ee
This gives
\be
Q\phi = c\phi\ ,\quad Q c = 0\ , \quad Q J_{\pm } = \pm c J_{\pm }\ ,
\ee
and importantly
\be
QJ_{3}(w) = \oint \frac{dz}{2\pi i} ~c(z) J_3(z)J_3(w) = \oint \frac{dz}{2\pi i} ~\frac{2\partial c(z)(z-w)}{(z-w)^2} = 2\partial c(w)\ .
\ee
Therefore both $J_3$ and $\partial c$ are not in $Q$-cohomology. The only gauge invariant operator that survives is
\be\label{gauge invariant sl2}
:\phi^2J_-:\ ,
\ee
which has the twisted spin 1, and a trivial OPE with itself. This operator can be identified with the boundary value of the bulk monopole operator $\phi^2 V_-$. The conformal dimension 1 space of the algebra may be viewed as the simplest example of an intermediate Lie algebra \eqref{intermediate lie algebra} with $\mathfrak{g}=\mathfrak{sl}_2$ \cite{landsberg2006sextonions}. 

In the topological $A$-twist, the operator \eqref{gauge invariant sl2} has twisted spin 2, and it is reasonable to expect that it corresponds (or at least contributes) to the stress-tensor of $M(5,3)$.

\section{Boundary vertex algebras for \texorpdfstring{$\CT_{\text{min}}$}{Tmin} coupled to \texorpdfstring{$L_1(\mathfrak{e}_8)$}{L1e8}}
 \label{sec: T1 e8}

We now consider the case where we take the boundary degrees of freedom to be $L_1(\mathfrak{e}_8)$, which is distinguished by the fact it defines a holomorphic CFT. By coupling it appropriately to $\overline\CT_{\text{min}}^A$, we will argue that the algebra of boundary operators realizes the minimal W-algebra $W_{-5}(\mathfrak{e}_8,f_\text{min})$, leading to a new level-rank type duality between $M(5,2)$ and $W_{-5}(\mathfrak{e}_8,f_\text{min})$.  In Section \ref{sec: HTB E7.5}, we discuss the boundary algebra of the $HT^B$-twisted theory, and argue that it is naturally identified with an intermediate vertex algebra, commonly denoted $(E_{7\frac12})_1$, which fills the ``missing hole" in the Deligne–Cvitanović exceptional series.


\subsection{Coupling \texorpdfstring{$L_1(\mathfrak{e}_8)$}{L1e8} to \texorpdfstring{$\overline\CT_{\text{min}}^A$}{TminA}} \label{sec: coupling Tmin e8}

Let us consider the ${\mathfrak u}(1)$ subalgebra of $\mathfrak{e}_8$ generated by $h_\theta = \theta^\vee$ (with $\theta$ the highest root), which defines the grading
\be
\mathfrak{e}_8 = {\mathfrak g}_{-2}\oplus {\mathfrak g}_{-1} \oplus {\mathfrak g}_{0} \oplus {\mathfrak g}_{1} \oplus {\mathfrak g}_{2}\ ,
\ee
where ${\mathfrak g}_0 = \mathfrak{g}_0' \oplus \mathbb{C}h_\theta$ with ${\mathfrak g}_0'=\mathfrak{e}_7$. As a representation of ${\mathfrak g}_0$, we can can decompose the adjoint representation of $\mathfrak{e}_8$ as 
\be
{\bf 248} = {\bf 1}_{-2} + {\bf 56}_{-1} + ({\bf 133} + {\bf 1})_0 + {\bf 56}_1 + {\bf 1}_2\ ,
\ee
where the subscript on each $\mathfrak{e}_7$ representation is its weight with respect to $h_\theta$.
We denote the corresponding generators of $L_1(\mathfrak{e}_8)$ by
\be\label{current decomposition}
 J^{\alpha}_{\mathfrak{e}_7}(z)\ ,\quad J_{h_\theta}(z)\ ,\quad J^i_{(\pm 1)}(z)\ ,\quad J_{(\pm 2)}(z)\ .
\ee

These boundary degrees of freedom are coupled to $\overline{\CT}_{\text{min}}$ with the Neumann boundary condition by identifying the subalgebra generated by $h_\theta$ with the current for the boundary ${\mathfrak u}(1)$ gauge symmetry. Since the corresponding current $J_{h_\theta}$ has level 2, its contribution precisely cancels the pure gauge anomaly of $\overline{\CT}_\text{min}$. The mixed gauge-global anomaly can likewise be canceled by assigning to the $L_1({\mathfrak e}_8)$ currents an $R_\nu$ charge equal to their charge under $\frac{1}{2}(1-\nu)h_\theta$. The contribution of the currents to the Neumann half-index is
\be
I_{2d}(q;z,\nu,\eta) =  \frac{1}{(q)^8_\infty}\sum_{n\in \mathbb{Z}^8} q^{\frac12 n^t C(E_8) n} z^{-n_1}[(-q^{1/2})^{\nu-1}\eta]^{-n_1/2}\ ,
\ee
where $C_8$ is the Cartan matrix of $\mathfrak{e}_8$. Our convention is
\be
C_8 = \begin{pmatrix}
 2 & -1 &  0 &  0 &  0 &  0 &  0 &  0 \\
-1 &  2 & -1 &  0 &  0 &  0 &  0 &  0 \\
 0 & -1 &  2 & -1 &  0 &  0 &  0 &  0 \\
 0 &  0 & -1 &  2 & -1 &  0 &  0 &  0 \\
 0 &  0 &  0 & -1 &  2 & -1 &  0 & -1 \\
 0 &  0 &  0 &  0 & -1 &  2 & -1 &  0 \\
 0 &  0 &  0 &  0 &  0 & -1 &  2 &  0 \\
 0 &  0 &  0 &  0 & -1 &  0 &  0 &  2
\end{pmatrix}.
\ee
One can check that the index transforms as desired under the large gauge transformations:
\be
I_{2d}(q;qz,\nu,\eta)  = q^{-1}(-q^{1/2})^{1-\nu} \eta^{-1}z^{-2} I_{2d}(q;z,\nu,\eta)\ .
\ee

For the topological $A$-twist, the half-index of coupled system is
\begin{equation}
\begin{split}\label{A-twist half index e8}
I_{\text{half}}(q;-1,1) &= (q)_\infty \oint \frac{dz}{2\pi i z} \frac{1}{(z;q)_\infty} I_{2d}(q;z,-1,1)\\
& = \frac{1}{(q)^7_\infty}\sum_{n_1\geq 0}\sum_{{(n_2,\cdots, n_8)}\in \mathbb{Z}^7}\frac{ q^{\frac12 n^t C_8 n} (-q^{1/2})^{n_1}}{(q)_{n_1}}\ .
\end{split}
\end{equation}

\subsection{Decomposition of \texorpdfstring{$L_1(\mathfrak{e}_8)$}{L1e8}}

Let $W^k(\mathfrak g, f_{\text{min}})$ be the universal minimal $W$-algebra of the Lie algebra $\mathfrak g$ at level $k$ and denote by $W_k(\mathfrak g, f_{\text{min}})$ its simple quotient. The Urod Theorem says that there is a  vertex algebra homomorphism
\[
\varphi: W^{k+n}(\mathfrak g, f_{\text{min}}) \rightarrow W_k(\mathfrak g, f_{\text{min}}) \otimes L_n(\mathfrak g)
\]
for any positive integer $n$. Here, the conformal vector of $L_n(\mathfrak g)$ has to be replaced by the Urod conformal vector (which in particular shifts the central charge by $-6n$).
Often this homomorphism factors through the simple quotient. This happens especially if the image of $\varphi$ is the coset   $\text{Com}(V, W_k(\mathfrak g, f_{\text{min}}) \otimes L_n(\mathfrak g))$ for some strongly rational vertex operator algebra $V$. We will only consider that instance. 

If we take $\mathfrak g = \mathfrak e_8$ and $k = -6$ and use $W_{-6}(\mathfrak e_8, f_{\text{min}}) \cong \mathbb C$ \cite{Adamovic:2016jjs} then there is a vertex algebra homomorphism $W^{-5}(\mathfrak e_8, f_{\text{min}}) \rightarrow L_1(\mathfrak e_8)$. Its image contains $L_1(\mathfrak e_7)$ as well as fields of conformal weight $3/2$ in the standard representation $\rho_{\omega_1}$ of $\mathfrak e_7$. 
We decompose
\begin{equation} \label{eq decomposition e8 1}
    \begin{split}
        L_1(\mathfrak e_8) &\cong L_1(\mathfrak e_7) \otimes L_1(\mathfrak{sl_2}) \oplus L_1(\omega_1) \otimes L_1(\omega) \\
        &= L_1(\mathfrak e_7) \otimes \left(M(5, 2) \otimes M(5, 3) \oplus  M^{5, 2}_{3, 1} \otimes    M^{5, 3}_{3, 1}\right) \oplus  \\
        &\qquad  L_1(\omega_1) \otimes \left(M^{5, 2}_{2, 1} \otimes    M^{5, 3}_{2, 1} \oplus  M^{5, 2}_{4, 1} \otimes    M^{5, 3}_{4, 1}  \right) \\
        &\cong  M(5, 2) \otimes \left(  L_1(\mathfrak e_7) \otimes M(5, 3) \oplus L_1(\omega_1) \otimes M^{5, 3}_{4, 1}  \right) \oplus \\
        &\qquad M^{5, 2}_{3, 1} \otimes \left(  L_1(\mathfrak e_7) \otimes M^{5, 3}_{3,1 } \oplus L_1(\omega_1) \otimes M^{5, 3}_{2, 1}  \right)\ .
    \end{split}
\end{equation}
We see that 
\be \label{com M52}
\text{Com}\left( M(5,2), L_1(\mathfrak e_8)\right) \cong  L_1(\mathfrak e_7) \otimes M(5, 3) \oplus L_1(\omega_1) \otimes M^{5, 3}_{4, 1} \ ,
\ee
which is exactly the simple minimal $W$-algebra $W_{-5}(\mathfrak e_8, f_{\text{min}})$ \cite{kawasetsu2018algebras}.
As it is a simple current extension of $L_1(\mathfrak e_7) \otimes M(5, 3)$  and the simple current 
$L_1(\omega_1) \otimes M^{5, 3}_{4, 1} $ has no fixed points
its modules can be all obtained via induction. One gets two local modules
\be
M_0 = L_1(\mathfrak e_7) \otimes M(5, 3) \oplus L_1(\omega_1) \otimes M^{5, 3}_{4, 1}, \qquad
M_1 = L_1(\mathfrak e_7) \otimes M^{5, 3}_{3,1 } \oplus L_1(\omega_1) \otimes M^{5, 3}_{2, 1} 
\ee
and two Ramond twisted modules
\be
M_2 = L_1(\mathfrak e_7) \otimes M^{5, 3}_{2,1 }  \oplus L_1(\omega_1) \otimes M^{5, 3}_{3, 1}, \qquad
M_3 = L_1(\mathfrak e_7) \otimes M^{5, 3}_{4,1 } \oplus L_1(\omega_1) \otimes M(5, 3).
\ee
The Urod conformal vector has zero mode $L_0 + h_0$ where $h$ is the element of the Cartan subalgebra that belongs to the same $\mathfrak{sl}_2$-triple as $f_{\text{min}}$. In this case it belongs to the weight $\omega_1 + \omega$ for the $\mathfrak e_7 \oplus \mathfrak{sl}_2$-subalgebra. 

The relation \eqref{eq decomposition e8 1} is realized via the interval reduction of $\CT_{\text{min}}^A$ with two distinct boundary conditions. On the right boundary, we impose the Dirichlet boundary condition which supports $M(5,2)$, while on the left boundary, we impose the Neumann boundary condition coupled to $L_1(\mathfrak{e}_8)$, albeit with reversed braidings, as described in Section \ref{sec: coupling Tmin e8}. This picture leads to the claim that the left boundary supports the simple minimal W-algebra $W_{-5}(\mathfrak e_8, f_\text{min})$. Since the module categories of the VOAs on the left and right boundaries are both realized as categories of bulk line operators, the two VOAs are naturally expected to have equivalent categories of modules, consistent with \eqref{eq decomposition e8 1} \cite{Creutzig:2017anl, Creutzig:2019psu}. The claim is further supported below by an explicit calculation of the (super)characters of its modules, which can be shown to agree with the Neumann half-index \eqref{A-twist half index e8}.

\subsection{Characters and supercharcters}

The relation between the characters with respect to the Urod conformal vector and the usual characters are
 \[
\text{ch}^U[L_1(\mathfrak{sl_2})](u, \tau) = \text{ch}[L_1(\omega)](u, \tau), \qquad 
\text{ch}^U[L_1(\omega)](u, \tau) = \text{ch}[L_1(\mathfrak{sl_2})](u, \tau).
\]
and 
\[
\text{ch}^U[L_1(\mathfrak{e}_7)](u, \tau) = \text{ch}[L_1(\omega_1)](u, \tau), \qquad 
\text{ch}^U[L_1(\omega_1)](u, \tau) = \text{ch}[L_1(\mathfrak{e}_7)](u, \tau).
\]
Thus
\begin{equation}
    \begin{split}
\text{ch}^U[M_0](u, \tau) &= \text{ch}[M_3](u, \tau), \qquad 
\text{ch}^U[M_1](u, \tau) = \text{ch}[M_2](u, \tau), \qquad  \\
\text{ch}^U[M_2](u, \tau) &= \text{ch}[M_1](u, \tau), \qquad 
\text{ch}^U[M_3](u, \tau) = \text{ch}[M_0](u, \tau).
 \end{split}
\end{equation}
The characters of $M_0$ and $M_1$ don't close under the modular $S$-transformation, but their supercharacters
\begin{equation}
    \begin{split}
\text{sch}[M_0] &= \text{ch}[L_1(\mathfrak e_7)] \text{ch}[M(5, 3)] - \text{ch}[L_1(\omega_1)] \text{ch}[M^{5, 3}_{4, 1}], \qquad \\
\text{sch}[M_1] &= \text{ch}[L_1(\omega_1)]\text{ch}[M^{5, 3}_{2, 1}] - \text{ch}[L_1(\mathfrak e_7)] \text{ch}[M^{5, 3}_{3, 1}]
 \end{split}
\end{equation}
do, namely the corresponding $S$-matrix is
\[
\sqrt{\frac{4}{5}}
\begin{pmatrix}
    \sin\left( \frac{2 \pi }{5} \right) & \sin\left( \frac{ \pi }{5} \right)\\[1mm]
    \sin\left( \frac{ \pi }{5} \right) &  - \sin\left( \frac{2 \pi }{5} \right) 
\end{pmatrix}
\]
and the Verlinde formula computes the superdimension of the fusion rules as required \cite{creutzig2024tensor}, especially
\[
\text{sch}[M_1] \times \text{sch}[M_1] = \text{sch}[M_0] - \text{sch}[M_1] = \text{sch}[M_1 \boxtimes M_1].
\]

One can verify that the $A$-twisted half-index \eqref{A-twist half index e8} exactly reproduces the supercharacter of the vacuum module $M_0$. To see this, we use
\be
\text{ch}(L_1(\mathfrak e_7))= q^{-7/24}\frac{1}{(q)^7_\infty}\sum_{n\in\mathbb{Z}^7} q^{\frac12 n  C(E_7) n^t}\ ,
\ee
\be
\text{ch}(L_1(w_1))= q^{11/24}\frac{1}{(q)^7_\infty}\sum_{n\in\mathbb{Z}^7} q^{\frac12 n  C(E_7) n^t-n_1}\ ,
\ee
and
\be
\text{ch}(M(5,3)) = q^{1/40} \sum_{n\geq 0} \frac{q^{n(n+1)}}{(q)_{2n}} \ ,\quad \text{ch}(M^{5,3}_{4,1}) = q^{-9/40} \sum_{n\geq 0} \frac{q^{(n+1)^2}}{(q)_{2n+1}}\ ,
\ee
\be
\text{ch}(M^{5,3}_{2,1}) = q^{-1/40} \sum_{n\geq 0} \frac{q^{n^2}}{(q)_{2n}} \ ,\quad \text{ch}(M^{5,3}_{3,1}) = q^{9/40} \sum_{n\geq 0} \frac{q^{n^2+n}}{(q)_{2n+1}}\ .
\ee
The $A$-twisted half-index \eqref{A-twist half index e8} decomposes into even and odd $n_1$ sectors:
\begin{equation}
    \begin{split}
I_{\text{half}}(q;-1,1) =&~ \frac{1}{(q)^7_\infty}\sum_{n_1\geq 0}\sum_{{(n_2,\cdots, n_8)}\in \mathbb{Z}^7}\frac{ q^{\frac12 n^t C_8 n} (-q^{1/2})^{n_1}}{(q)_{n_1}} \\
=&~ \frac{1}{(q)_\infty^7}\sum_{\ell\geq 0}\sum_{{\bf n'}\in \mathbb{Z}^7} \frac{q^{\frac12 {\bf n'}^t C(E_7) {\bf n'}-2\ell n_2+4\ell^2+\ell}}{(q)_{2\ell}} \\
&- \frac{1}{(q)_\infty^7}\sum_{\ell\geq 0}\sum_{{\bf n'}\in \mathbb{Z}^7} \frac{q^{\frac12 {\bf n'}^t C(E_7) {\bf n'}-(2\ell+1) n_2+(2\ell+1)^2+\ell+\frac12}}{(q)_{2\ell+1}}\ .
\end{split}
\end{equation}
Shifting ${\bf n'} \rightarrow {\bf n'}+2\ell w_1$,
\begin{equation}
\begin{split}
I_{\text{half}}(q;-1,1)  =&~\frac{1}{(q)_\infty^7} \sum_{\ell\geq 0}\sum_{{\bf n'}\in \mathbb{Z}^7} \frac{q^{\frac12 {\bf n'}^t C(E_7) {\bf n'}+\ell^2+\ell}}{(q)_{2\ell}} \\
&-  \frac{q^{1/2}}{(q)_\infty^7} \sum_{\ell\geq 0}\sum_{{\bf n'}\in \mathbb{Z}^7} \frac{q^{\frac12 {\bf n'}^t C(E_7) {\bf n'}-n_2+(\ell+1)^{2}}}{(q)_{2\ell+1}}
\end{split}
\end{equation}
which indeed reproduces sch$[M_0]$, up to a modular anomaly prefactor.

We can repeat the same analysis with $W_1$, the Wilson line of gauge charge 1, inserted.  The half-index in the presence of $W_1$ reads
\begin{equation}
\begin{split}
I_{\text{half}}[W_1](q;-1,1) = &~ (q)_\infty \oint \frac{dz}{2\pi i z} \frac{1}{(z;q)_{\infty}}  I_{2d}(q;z,-1,1 )\cdot z^{-1}\\
=&~ \frac{1}{(q)^7_\infty}\sum_{n_1\geq -1}\sum_{{(n_2,\cdots, n_8)}\in \mathbb{Z}^7}\frac{ q^{\frac12 n^t C(E_8) n} (-q^{1/2})^{n_1}}{(q)_{n_1+1}}\ ,
\end{split}
\end{equation}
which can be decomposed into
\begin{equation}
\begin{split}
I_{\text{half}}[W_1](q;-1,1) =&~\frac{1}{(q)_\infty^7}\sum_{\ell\geq 0}\sum_{{\bf n'}\in \mathbb{Z}^7} \frac{q^{\frac12 {\bf n'}^t C(E_7) {\bf n'}-2\ell n_2+4\ell^2+\ell}}{(q)_{2\ell+1}} \\
&- \frac{1}{(q)_\infty^7}\sum_{\ell\geq -1}\sum_{{\bf n'}\in \mathbb{Z}^7} \frac{q^{\frac12 {\bf n'}^t C(E_7) {\bf n'}-(2\ell+1) n_2+(2\ell+1)^2+\ell+\frac12}}{(q)_{2\ell+2}}\ .
\end{split}
\end{equation}
Shifting ${\bf n'} \rightarrow {\bf n'}+2\ell w_1$, 
\begin{equation}
\begin{split}
I_{\text{half}}[W_1](q;-1,1) =& ~\frac{1}{(q)_\infty^7} \sum_{\ell\geq 0}\sum_{{\bf n'}\in \mathbb{Z}^7} \frac{q^{\frac12 {\bf n'}^t C(E_7) {\bf n'}+\ell^2+\ell}}{(q)_{2\ell+1}}  \\
&- \frac{q^{1/2}}{(q)_\infty^7}\sum_{\ell\geq 0}\sum_{{\bf n'}\in \mathbb{Z}^7} \frac{q^{\frac12 {\bf n'}^t C(E_7) {\bf n'}-n_2+\ell^2}}{(q)_{2\ell}}\ ,
\end{split}
\end{equation}
which again reproduces sch$[M_1]$ up to a modular anomaly prefactor.

\subsection{\texorpdfstring{$HT^B$}{HTB}-twist and \texorpdfstring{$(E_{7\scriptsize{\frac12}})_1$}{E71/21}} 
\label{sec: HTB E7.5}

As in Section \ref{sec: HTB sl2}, one may consider the $HT^B$-twisted theory coupled to $L_1(\mathfrak{e}_8)$. The Neumann half-index in the limit $\nu=\eta=1$ becomes
\begin{equation}
\begin{split}\label{E7.5 character}
I_{\text{half}}(q;1,1) =&~ (q)_\infty \oint \frac{dz}{2\pi i z} \frac{1}{(z;q)_\infty} I_{2d}(q;z,1,1)\\
= &~\frac{1}{(q)^7_\infty}\sum_{n_1\geq 0}\sum_{{(n_2,\cdots, n_8)}\in \mathbb{Z}^7}\frac{ q^{\frac12 n^t C_8 n} }{(q)_{n_1}}\ , \\
=&~1+ 190q + 2831q^2 + 22306q^3+\cdots\ ,
\end{split}
\end{equation}
where the contribution from $L_1(\mathfrak{e}_8)$ is now taken with respect to the standard conformal vector, corresponding to the $HT^B$-twisted spin. 
Notice that the resulting $q$-series coincides precisely with the "missing hole" appearing in the MMS classification of 2d rational conformal field theories with two characters \cite{Mathur:1988na}. This $q$-series is often associated with a putative vertex algebra denoted by $(E_{7\frac12})_1$,
where $E_{7\frac12}$ is the intermediate Lie algebra \eqref{intermediate lie algebra} corresponding to $\mathfrak{g}=\mathfrak{e}_8$ \cite{landsberg2006sextonions,kawasetsu2014intermediate}.

For the $HT^B$-twisted theory with the standard Neumann boundary condition, the boundary operator algebra can be calculated explicitly. The $Q$-cohomology contains three classes of gauge-invariant boundary operators,
\be
 J^{\alpha}_{\mathfrak{e}_7}\ ,\quad :\phi J^i_{(-1)}:\ ,\quad :\phi^2 J_{(- 2)}:\ ,
\ee
all of which have twisted spin one. They form an $\mathfrak{e}_7$ representation
\be
{\bf 190} = {\bf 133}+{\bf 56}+{\bf 1}\ ,
\ee
which is isomorphic to $\mathfrak{e}_7 \oplus \mathfrak{g}_1 \oplus \mathfrak{g}_2$, the intermediate Lie algebra corresponding to $\mathfrak{g}=\mathfrak{e}_8$. 
Let $A^i = \phi J^i_{(-1)}$ and $B =\phi^2 J_{(-2)}$. The singular OPEs are
\begin{equation}
\begin{split}
& J^{\alpha}_{\mathfrak{e}_7}(z)  J^{\beta}_{\mathfrak{e}_7}(w) \sim \frac{\kappa^{\alpha\beta}}{(z-w)^2} + \frac{f^{\alpha\beta}_\gamma J^\gamma_{\mathfrak{e}_7}(w)}{z-w}\ ,  \\ 
&J^{\alpha}_{\mathfrak{e}_7}(z)  A^i(w) \sim  \frac{{f^{\alpha i}}_j A^j(w)}{z-w}\ , \quad  A^i(z)A^j(w) \sim \frac{f^{ij}_{-\theta} B(w)}{z-w} 
\end{split}
\end{equation}
and all the other OPEs are regular.  Notice that $B(z)$ have regular OPEs with all the operators, which is consistent with the expectation that the operator $B(z)$ is the boundary value of the bulk monopole operator $\phi^2 V_-$.

\section{Boundary vertex algebras for \texorpdfstring{$\CT_2$}{T2} coupled to \texorpdfstring{$L_1(\mathfrak{e}_8)$}{L1e8}} \label{sec: T2}

The minimal SCFT $\CT_{\text{min}}$ is the simplest member of a one-parameter family of rank-zero $\CN=4$ SCFTs $\CT_r$, with $\CT_1 = \CT_{\text{min}}$, whose topological $A$-twist admits a boundary condition that supports the Virasoro minimal model $M(2r+3,2)$ \cite{Gang:2023rei}. In this section, we consider coupling the 2d degrees of freedom $V_{2d}=L_1(\mathfrak{e}_8)$ to the next simplest theory in the family, $\CT_2$. This theory is expected to have a UV realization in terms of a $\CN=2$ $U(1)^2$ CS-matter theory description, where each $U(1)$ gauge group couples to a single chiral multiplet of charge 1. The Chern-Simons level matrix is $\kappa_{ij} =2\text{min}(i,j)$, with $i,j=1,2$. Upon turning on the superpotential deformation
\be
W = V_{(2,-1)}\ ,
\ee
where $V_{(m,n)}$ denotes the bare monopole operator with fluxes $(m,n)\in \mathbb{Z}^2$, the theory is expected to flow to the SCFT, $\CT_2$, whose Coulomb and Higgs branches are both zero-dimensional. The CS-matter theory has global symmetry $U(1)_R\times U(1)_S$, where $U(1)_S$ can be identified with the unbroken global symmetry $U(1)_{1}^{\text{top}} + 2U(1)_{2}^{\text{top}}$, where $U(1)_{i}^{\text{top}}$ is the topological symmetry associated with the $i$-th gauge node. This global symmetry enhances to the full R-symmetry group $SO(4)$ in the IR SCFT. It is argued in \cite{Gang:2023rei,Ferrari:2023fez} that the deformed Dirichlet boundary condition for $\CT_2^A$ realizes $M(7,2)$.

We consider the topological A-twist of the parity reversed theory, $\overline{\CT}^A_2$, with a Neumann boundary condition. The bulk theory induces a boundary gauge anomaly of the form 
\be\label{anomaly T2}
-2{\bf f}_1^2 - 4{\bf f}_2^2 -4 {\bf f}_1{\bf f}_2 - 2({\bf s}-{\bf r})({\bf f}_1+2{\bf f}_2)\ ,
\ee 
where ${\bf s}$ is the background field strength for the $U(1)_S$ symmetry.
We will show that this anomaly can be completely canceled by coupling boundary degrees of freedom $L_1(\mathfrak{e}_8)$ in a specific way and then argue that the algebra of boundary operators realizes a rational and $C_2$-cofinite VOA that is level-rank dual to $M(7,2)$. In Section \ref{sec: X1}, we turn to the boundary algebra of the $HT^B$-twisted theory and argue that it can be identified with an intermediate algebra, denoted $(X_1)_1$, which appears in \cite{Mkrtchyan:2012es,Lee:2024fxa,Kim:2024dxu}.

\subsection{Coupling \texorpdfstring{$L_1(\mathfrak{e}_8)$}{L1e8} to \texorpdfstring{$\overline\CT_2^A$}{T2A}} \label{sec: coupling e8 to T2}

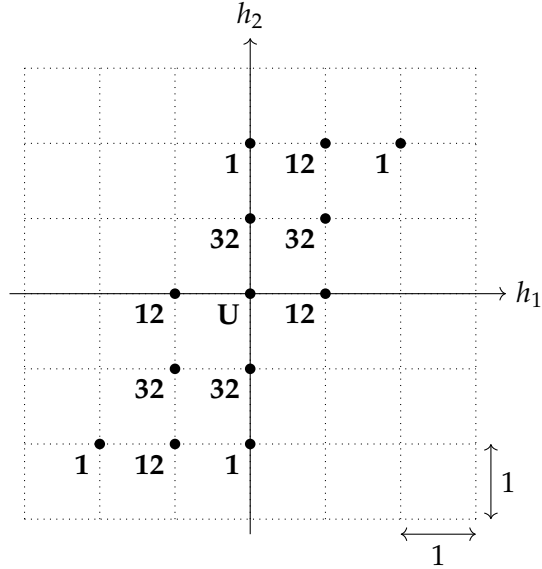
\begin{figure}[ht]
  \centering
\begin{tikzpicture}[scale=1] \label{fig: decomposition X1}
  \draw[step=1cm, dotted] (-3,-3) grid (3,3);

  \draw[->] (-3.2,0) -- (3.4,0) node[right] {$h_1$};

  \draw[->] (0,-3.2) -- (0,3.4) node[above] {$h_2$};

  \fill (0,-2) circle (2pt) node[below left] {$\mathbf{1}$};
  \fill (0,-1) circle (2pt) node[below left] {$\mathbf{32}$};
  \fill (0,0) circle (2pt) node[below left] {${\mathbf U}$};
  \fill (0,1) circle (2pt) node[below left] {$\mathbf{32}$};
  \fill (0,2) circle (2pt) node[below left] {$\mathbf{1}$};
  \fill (1,0) circle (2pt) node[below left] {$\mathbf{12}$};
  \fill (1,1) circle (2pt) node[below left] {$\mathbf{32}$};
  \fill (1,2) circle (2pt) node[below left] {$\mathbf{12}$};
  \fill (2,2) circle (2pt) node[below left] {$\mathbf{1}$};
  \fill (-1,0) circle (2pt) node[below left] {$\mathbf{12}$};
  \fill (-1,-1) circle (2pt) node[below left] {$\mathbf{32}$};
  \fill (-1,-2) circle (2pt) node[below left] {$\mathbf{12}$};
  \fill (-2,-2) circle (2pt) node[below left] {$\mathbf{1}$};
  \draw[<->] (2,-3.2) -- (3,-3.2)
    node[midway, below] {$1$};
  \draw[<->] (3.2,-3) -- (3.2,-2)
    node[midway, right] {$1$};
\end{tikzpicture}
\caption{A decomposition of $\mathfrak{e}_8$. The lattice points are labeled by their $\mathfrak{so}_{12}$ representations. We have ${\bf U} = \mathfrak{g}_{0,0} = {\bf 66} + {\bf 1} + {\bf 1}$. }
\end{figure}

Let us consider the $\mathfrak{u}(1)\oplus \mathfrak{u}(1)$ subalgebra of $\mathfrak{e}_8$ generated by $h_1 = \theta^\vee = \omega_1^\vee$ and $h_2 = \omega_7^\vee$, where $w_i$ denotes the $i$-th fundamental weight.\footnote{In the simple root basis, $w_1 = (2,3,4,5,6,4,2,3)\ , w_7 = (2,4,6,8,10,7,4,5)$.} The Lie algebra $\mathfrak{e}_8$ admits a decomposition according to the charges under $h_1$ and $h_2$,
\be
\mathfrak{e}_8 = \bigoplus_{i,j=-2}^2{\mathfrak g}_{i,j}\ .
\ee
Here $\mathfrak{g}_{0,0} \simeq \mathfrak{so}_{12} \oplus \mathbb{C}h_1 \oplus \mathbb{C}h_2$. The decomposition as a $\mathfrak{g}_{0,0}$-representation is summarized in Figure \ref{fig: decomposition X1}.

We denote the corresponding currents by 
\be
J^\alpha_{\mathfrak{so}_{12}},~  J_{h_1},~ J_{h_2},~J_{\pm (0,2)}~J^i_{\pm (1,0)},~J^i_{\pm (1, 2)},~J^a_{\pm(0,1)},~J^a_{\pm(1, 1)},~J_{\pm ( 2, 2)}\ .
\ee
The two $\mathfrak{u}(1)$ currents have the OPE
\be
J_{h_i}(z)J_{h_j}(w) = \frac{\kappa_{ij}}{(z-w)^2}\ , \qquad i,j=1,2\ ,
\ee
where $\kappa_{ij} = 2\text{min}(i,j)$ is the CS level of the bulk gauge theory, and therefore it cancels the pure gauge anomalies from the bulk $\overline\CT_2$. As in the previous example, the mixed gauge-global anomaly can be canceled by assigning appropriate $R$-charges to the $L_1(\mathfrak{e}_8)$ currents. We claim that this is achieved by assigning each current an $R_\nu$-charge equal to its charge under $\frac12(1-\nu)h_2$. With this assignment, the contribution of the currents to the Neumann half-index is
\be\label{character of e8 current T2}
I_{2d}(q;z_1,z_2,\nu,\eta) = \frac{1}{(q)^8_\infty}\sum_{n\in \mathbb{Z}^8} q^{\frac12 n^t C(E_8) n} z_1^{-n_1}z_2^{-n_7}[(-q^{1/2})^{\nu-1}\eta]^{-n_7/2}\ .
\ee
Indeed, one can check that under the large gauge transformation $z_1\rightarrow z_1 q$ and $z_2\rightarrow z_2 q$, it transforms as
\begin{equation}
\begin{split}
&I(q;z_1q,z_2,\nu,\eta) = q^{-1}[(-q^{1/2})^{\nu-1}\eta]^{-1} z_1^{-2}z_2^{-2}I(q;z_1,z_2,\nu,\eta)\ ,\\
&I(q;z_1,z_2q,\nu,\eta) = q^{-2}[(-q^{1/2})^{\nu-1}\eta]^{-2} z_1^{-2}z_2^{-4}I(q;z_1,z_2,\nu,\eta)\ ,
\end{split}
\end{equation}
which implies that the contribution of these degrees of freedom exactly cancels the gauge anomaly \eqref{anomaly T2}.

For the topological $A$-twist, the Neumann half-index of the coupled system is
\begin{equation}\label{Neumann character T2}
\begin{split}
I_{\text{half}}(q;-1,1) &= (q)_\infty^2 \oint \frac{dz_1}{2\pi i z_1}\frac{dz_2}{2\pi i z_2} \frac{1}{(z_1;q)_\infty(z_2;q)_\infty} I_{2d}(q;z_1,z_2,-1,1) \\
&= \frac{1}{(q)^6_\infty}\sum_{n_1,n_7\geq 0}\sum_{{(n_2,\cdots, n_6,n_8)}\in \mathbb{Z}^6}\frac{ q^{\frac12 n^t C(E_8) n} (-q^{1/2})^{n_7}}{(q)_{n_1}(q)_{n_7}}\\
&=1+78q - 64 q^{3/2} + 898 q^2 - 896 q^{5/2} + 6072 q^3 -\cdots\ . 
\end{split}
\end{equation}

\subsection{Decomposition of \texorpdfstring{$L_1(\mathfrak{e}_8)$}{L1e8}} We consider the decomposition
\be
L_1(\mathfrak{e}_8) = L_1(\mathfrak{e}_7)\otimes L_1(\mathfrak{sl}_2) \oplus L_1^{\mathfrak{e}_7}(\omega_1)\otimes L_1^{\mathfrak{sl}_2}(\omega)\ ,
\ee 
followed by a further decomposition
\begin{equation}
\begin{split}
L_1(\mathfrak{e}_7) &= L_1(\mathfrak{so}_{12})\otimes L_1(\mathfrak{sl}_2) \oplus L_1^{\mathfrak{so}_{12}}(\omega_6)\otimes L_1^{\mathfrak{sl}_2}(\omega)\ ,\\ 
L_1^{\mathfrak{e}_7}(\omega_1) &= L_1^{\mathfrak{so}_{12}}(\omega_5)\otimes L_1(\mathfrak{sl}_2) \oplus L_1^{\mathfrak{so}_{12}}(\omega_1)\otimes L_1^{\mathfrak{sl}_2}(\omega)\ ,
\end{split}
\end{equation}
which gives
\begin{equation}
\begin{split}\label{e8 twisted}
L_1(\mathfrak{e}_8) &= \left[L_1(\mathfrak{so}_{12})\otimes L_1(\mathfrak{sl}_2) \oplus L_1^{\mathfrak{so}_{12}}(\omega_6)\otimes L_1^{\mathfrak{sl}_2}(\omega)\right]\otimes L_1(\mathfrak{sl}_2) \\
&\oplus \left[ L_1^{\mathfrak{so}_{12}}(\omega_5)\otimes L_1(\mathfrak{sl}_2) \oplus L_1^{\mathfrak{so}_{12}}(\omega_1)\otimes L_1^{\mathfrak{sl}_2}(\omega)\right]\otimes L_1^{\mathfrak{sl}_2}(\omega)\ .
\end{split}
\end{equation}
If we choose the conformal vector of the two copies of $L_1(\mathfrak{sl}_2)$ and $L_1^{\mathfrak{sl}_2}(\omega)$ above to be the Urod conformal vector, the character of \eqref{e8 twisted} precisely reproduces the contribution of the the boundary degrees of freedom \eqref{character of e8 current T2}, in the $A$-twist limit $\nu=-\eta = -1$. Using the Urod-GKO coset decomposition for Virasoro 
\begin{equation}
    \begin{split}
M^{u, v}_{r, 1} \otimes L_1(\mathfrak{sl_2}) &\cong \bigoplus_{\substack{s =1 \\ s + r \ \text{even}}}^{u+v-1} M^{u+v, v}_{s, 1} \otimes M^{u+v, u}_{s, r} \\ 
M^{u, v}_{r, 1} \otimes L_1(\omega) &\cong \bigoplus_{\substack{s =1 \\ s + r \ \text{odd}}}^{u+v-1} M^{u+v, v}_{s, 1} \otimes M^{u+v, u}_{s, r} \ ,
  \end{split}
\end{equation}
we arrive at
\be
L_1(\mathfrak{e}_8) = \bigoplus_{t=1}^3 M_{t,1}^{7,2}\otimes C_t\ ,
\ee
where
\begin{equation}
\begin{split}\label{C1}
C_1=&~L_1(\mathfrak{so}_{12})\otimes \left[M(5,3) \otimes M(7,5) \oplus M^{5,3}_{3,1}\otimes M^{7,5}_{1,3} \right] \\
&\oplus L_1^{\mathfrak{so}_{12}}(\omega_6) \left[ M(5,3)\otimes M^{7,5}_{1,4}\oplus M^{5,3}_{3,1}\otimes M^{7,5}_{1,2}\right] \\
&\oplus~ L_1^{\mathfrak{so}_{12}}(\omega_1) \otimes \left[ M^{5,3}_{2,1}\otimes M^{7,5}_{1,2} \oplus M^{5,3}_{4,1}\otimes M^{7,5}_{1,4}\right] \\
&\oplus L_1^{\mathfrak{so}_{12}}(\omega_5) \left[ M^{5,3}_{2,1} \otimes M^{7,5}_{1,3} \oplus M^{5,3}_{4,1}\otimes M(7,5)\right]\ ,
\end{split}
\end{equation}
\begin{equation}
\begin{split}
C_2=&~L_1(\mathfrak{so}_{12})\otimes \left[M(5,3) \otimes M^{7,5}_{5,1} \oplus M^{5,3}_{3,1}\otimes M^{7,5}_{5,3} \right] \\
&\oplus L_1^{\mathfrak{so}_{12}}(\omega_6) \left[ M(5,3)\otimes M^{7,5}_{2,1}\oplus M^{5,3}_{3,1}\otimes M^{7,5}_{2,3}\right] \\
&\oplus~ L_1^{\mathfrak{so}_{12}}(\omega_1) \otimes \left[ M^{5,3}_{2,1}\otimes M^{7,5}_{5,2} \oplus M^{5,3}_{4,1}\otimes M^{7,5}_{5,4}\right] \\
& \oplus L_1^{\mathfrak{so}_{12}}(\omega_5) \left[ M^{5,3}_{2,1} \otimes M^{7,5}_{2,2} \oplus M^{5,3}_{4,1}\otimes M^{7,5}_{2,4}\right]\ ,
\end{split}
\end{equation}
and
\begin{equation}
\begin{split}
C_3=&~L_1(\mathfrak{so}_{12})\otimes \left[M(5,3) \otimes M^{7,5}_{3,1} \oplus M^{5,3}_{3,1}\otimes M^{7,5}_{3,3} \right] \\
& \oplus L_1^{\mathfrak{so}_{12}}(\omega_6) \left[ M(5,3)\otimes M^{7,5}_{4,1}\oplus M^{5,3}_{3,1}\otimes M^{7,5}_{4,3}\right] \\
&\oplus~ L_1^{\mathfrak{so}_{12}}(\omega_1) \otimes \left[ M^{5,3}_{2,1}\otimes M^{7,5}_{3,2} \oplus M^{5,3}_{4,1}\otimes M^{7,5}_{3,4}\right] \\
&\oplus L_1^{\mathfrak{so}_{12}}(\omega_5) \left[ M^{5,3}_{2,1} \otimes M^{7,5}_{4,2} \oplus M^{5,3}_{4,1}\otimes M^{7,5}_{4,4}\right]\ .
\end{split}
\end{equation}
From this we have
\be
\text{Com}(M(7,2),L_1(\mathfrak{e}_8))= C_1\ .
\ee


As in the previous section, we consider the interval reduction of $\CT_2^A$ with two
distinct boundary conditions. On the right boundary, we impose the exceptional Dirichlet boundary conditions of \cite{Gang:2023rei} which supports the VOA $M(7,2)$, while on the left boundary, we impose the Neumann boundary condition coupled to $L_1(\mathfrak{e}_8)$, as described in Section \ref{sec: coupling e8 to T2}. The VOAs supported on the left and right boundaries are expected to have (braid-reversed) equivalent categories of modules. In particular, it is natural to expect that the VOA on the left boundary of $\CT_2^A$ (or, equivalently, the VOA on the right boundary of $\overline{\CT}^A_2$) can be realized as the coset of $L_1(\mathfrak{e}_8)$ by $M(7,2)$.

%

\subsection{Characters and supercharacters} The above claim is strongly supported by a comparison of characters. We find that the first few coefficients of \eqref{Neumann character T2} agree with those of the supercharacter of $C_1$ defined in \eqref{C1}, at least up to $q^4$.

The bulk TFT has three simple lines, $1, W_{(1,1)}, W_{(1,2)}$, where the latter two lines correspond to the Wilson lines of charge $(1,1)$ and $(1,2)$ under the $U(1)^2$ gauge group \cite{Gang:2023rei}, respectively. Inserting these lines, the half-indices are
\begin{equation}
\begin{split}
I_{\text{half}}[W_{(1,1)}](q;-1,1) =&~ (q)_\infty^2 \oint \frac{dz_1}{2\pi i z_1}\frac{dz_2}{2\pi i z_2} \frac{1}{(z_1;q)_\infty(z_2;q)_\infty} I_{2d}(q;z_1,z_2,-1,1)z_1^{-1}z_2^{-1} \\
=&~ \frac{1}{(q)^6_\infty}\sum_{n_1,n_7\geq -1}\sum_{{(n_2,\cdots, n_6,n_8)}\in \mathbb{Z}^6}\frac{ q^{\frac12 n^t C(E_8) n} (-q^{1/2})^{n_7}}{(q)_{n_1+1}(q)_{n_7+1}}
\end{split}
\end{equation}
and
\begin{equation}
\begin{split}
I_{\text{half}}[W_{(1,2)}](q;-1,1) =&~ (q)_\infty^2 \oint \frac{dz_1}{2\pi i z_1}\frac{dz_2}{2\pi i z_2} \frac{1}{(z_1;q)_\infty(z_2;q)_\infty} I_{2d}(q;z_1,z_2,-1,1)z_1^{-1}z_2^{-2} \\
=&~ \frac{1}{(q)^6_\infty}\sum_{\substack{n_1\geq -1\\n_7\geq -2}}\sum_{{(n_2,\cdots, n_6,n_8)}\in \mathbb{Z}^6}\frac{ q^{\frac12 n^t C(E_8) n} (-q^{1/2})^{n_7}}{(q)_{n_1+1}(q)_{n_7+2}}\ .
\end{split}
\end{equation}
We check that the first few coefficients of these $q$-series agree with the supercharacters of the modules $C_2$ and $C_3$ respectively.

\subsection{\texorpdfstring{$HT^B$}{HTB}-twist and \texorpdfstring{$(X_1)_1$}{X11}} \label{sec: X1} We now consider the $HT^B$-twisted $\overline\CT_2$ theory with a Neumann boundary condition coupled to boundary degrees of freedom given by $L_1(\mathfrak{e}_8)$. The above Neumann half-index in the limit $\nu=\eta=1$ reads
\begin{equation}
\begin{split}
I_{\text{half}}(q;1,1)=&~ (q)_\infty^2 \oint \frac{dz_1}{2\pi i z_1}\frac{dz_2}{2\pi i z_2}\frac{1}{(z_1;q)_\infty(z_2;q)_\infty} I_{2d}(q;z_1,z_2,1,1)\\
=&~\frac{1}{(q)^6_\infty}\sum_{n_1,n_7\geq 0}\sum_{{(n_2,\cdots, n_6,n_8)}\in \mathbb{Z}^6}\frac{ q^{\frac12 n^t C(E_8) n} }{(q)_{n_1}(q)_{n_7}}\ . \\\nonumber
=&~1 + 156 q + 2236 q^2 + 17056 q^3  +\cdots\ .
\end{split}
\end{equation}
This coincides with one of the modular invariant characters of the intermediate algebra $(X_1)_1$, between $D_6$ and $E_8$.

The gauge invariant boundary operators that survive the $Q$-cohomology are
\begin{equation}
\begin{split}
J^\alpha_{\mathfrak{so}_{12}}\ ,\quad :\phi_1J^i_{(-1,0)}:\ ,\quad :\phi_1\phi_2^2J^i_{(-1,-2)}:\ , \quad :\phi_2J^a_{(0,-1)}:\ ,\\
\quad :\phi_1^2\phi_2J^a_{(-1,-1)}:\ , \quad :\phi_2^2J_{(0,-2)}:\ ,\quad :\phi_1^2\phi_2^2J_{(-2,-2)}:\ ,
\end{split}
\end{equation}
all of which have twisted spin one. They form an $\mathfrak{so}_{12}$-representation
\be
{\bf 156} = {\bf 66} + {\bf 12} + {\bf 12} +  {\bf 32}+ {\bf 32} + {\bf 1}  +{\bf 1}\ ,
\ee
which is isomorphic to a non-reductive Lie algebra 
\be
\mathfrak{so}_{12}\oplus \bigoplus_{i+j>0} \mathfrak{g}_{i,j}\ ,
\ee
inside $\mathfrak{e}_8$. The OPEs among them are directly inherited from $L_1(\mathfrak{so}_{12})$. In particular, the operator $\phi_1^2\phi_2^2J_{(-2,-2)}$ has regular OPEs with everything. This operator is identified with the boundary value of the quarter-BPS bulk monopole operator $\phi_1^2\phi_2^2V_{(-1,0)}$, which is part of the extra supercurrent multiplet \cite{Gang:2023rei}.

\section{W-algebras of the Deligne–Cvitanovi\'{c} Exceptional series} \label{sec: DC}

More generally, the boundary gauge anomalies of $\overline{\CT}^A_{\text{min}}$ can be canceled by taking boundary degrees of freedom corresponding to $L_1(\mathfrak{g})$, where $\mathfrak{g}$ is a Lie algebra in the Deligne–Cvitanovi\'{c} (DC) exceptional series. As discussed in Section \ref{sec: coupling sl2}, this is realized by considering the boundary algebra of the coupled system $\overline{\CT}^A_{\text{min}}\times T_{\mathfrak{g}}$, where $T_{\mathfrak{g}}$ is the level-1 Chern-Simons theory with gauge group $G$, taken to be the simply connected group with Lie algebra $\mathfrak{g}$. 

\subsection{Coupling \texorpdfstring{$L_1(\mathfrak{g})$}{L1g} to \texorpdfstring{$\overline \CT_\text{min}^A$}{TminA}} 

Let $\chi_{\mathfrak g}(q;\{y_i\})$ be the character of $L_1(\mathfrak{g})$ with the standard conformal vector. The contribution of the above boundary degrees of freedom is then
\be
I^{\mathfrak{g}}_{2d}(q;z,\nu,\eta) = \chi_\mathfrak{g}\left(q;\left\{z^{a_i^\vee}[(-q^{1/2})^{\nu-1}\eta]^{a_i^\vee/2}\right\}\right)\ ,
\ee
where $a_i^\vee$ are the comarks, defined by $\theta^\vee = \sum_i a_i^\vee \alpha^\vee_i$. The shift of the Jacobi variable $z$ is due to the mixed Chern-Simons coupling \eqref{mixed gauge global}.
When $L_1(\mathfrak{g})$ is a lattice VOA, namely, for $\mathfrak{g}=A_1, A_2, D_4, E_6, E_7, E_8$, the contribution can be written in the form,
\be
I^{\mathfrak{g}}_{2d}(q;z,\nu,\eta) =  \frac{1}{(q)^{\text{rk}(\mathfrak{g})}_\infty}\sum_{n\in \mathbb{Z}^{\text{rk}(\mathfrak{g})}} q^{\frac12 n^t C(\mathfrak{g}) n} z^{-a_i^\vee C(\mathfrak{g})_{ij} n_j}[(-q^{1/2})^{\nu-1}\eta]^{-a_i^\vee C(\mathfrak{g})_{ij} n_j/2}\ ,
\ee
where $C(\mathfrak{g})$ is the Cartan matrix of $\mathfrak{g}$. For these examples, the Neumann half-index can be calculated explicitly. Let $n_{\theta^\vee} =\sum_{i,j}a_i^\vee C(\mathfrak{g})_{ij} n_j$. Then

\begin{equation}
\begin{split}\label{half index L1g}
I^{\mathfrak{g}}_{\text{half}}(q;\nu,\eta) =&~ (q)_\infty \oint \frac{dz}{2\pi i z} \frac{1}{(z;q)_\infty}I^{\mathfrak{g}}_{2d}(q;z,\nu,\eta) \\
=&~ \frac{1}{(q)^{\text{rk}(\mathfrak{g})-1}_\infty}\sum_{\substack{n\in \mathbb{Z}^{\text{rk}(\mathfrak{g})}\\n_{\theta^\vee}\geq 0}} \frac{1}{(q)_{n_{\theta^{\vee}}}} q^{\frac12 n^t C(\mathfrak{g}) n} [(-q^{1/2})^{\nu-1}\eta]^{-n_{\theta^\vee}/2}\ .
\end{split}
\end{equation}
In the $A$-twist limit, we have
\be
I^{\mathfrak{g}}_{\text{half}}(q;-1,1) =\frac{1}{(q)^{\text{rk}(\mathfrak{g})-1}_\infty}\sum_{\substack{n\in \mathbb{Z}^{\text{rk}(\mathfrak{g})}\\n_{\theta^\vee}\geq 0}} \frac{1}{(q)_{n_{\theta^{\vee}}}} q^{\frac12 n^t C(\mathfrak{g}) n} (-q^{1/2})^{n_{\theta^\vee}}\ .
\ee

\subsection{Decomposition of \texorpdfstring{$L_1(\mathfrak{g})$}{L1g}}
Let us recall some peculiarities of the Deligne–Cvitanovi\'{c}  exceptional series. 
Let $\mathfrak{g}$ be a Lie algebra of the DC series and let $h^\vee$ be its dual Coxeter number. Then the dimension of $\mathfrak g$ satisfies
\[
\text{dim}(\mathfrak{g}) = \frac{2(h^\vee+1)(5h^\vee-6)}{h^\vee + 6}.
\]
This has a few combinatorical consequences for the associated vertex algebras. The central charge of the minimal $W$-algebra of $\mathfrak{g}$ at level $k$ is \cite{KacWak}
\be
c(k) = \frac{k\text{dim}\mathfrak{g}}{k+h^\vee} - 6k + h^\vee -4\ .
\ee
Moreover in the case that $\mathfrak g$ is in the DC-series the level of the affine vertex subalgebra $V^{k^\sharp}(\mathfrak g^\sharp)$ of 
$W^{k}(\mathfrak{g}, f_\text{min})$ is 
\[
k^\sharp = k + \frac{h^\vee}{6} +1.
\]
This level is zero for $k = -\frac{h^\vee}{6} -1$ and one verifies that the central charge of $W_{k}(\mathfrak{g}, f_\text{min})$ is also zero at this level. It then turns out that this is a collapsing level, that is $W_{k}(\mathfrak{g}, f_\text{min}) \cong \mathbb C$ is trivial \cite{Adamovic:2016jjs, Adamovic:2025ilx}. 

The next coincidence appears when considering the coset $\text{Com}(V^{k+1}(\mathfrak g), V^k(\mathfrak g) \otimes L_1(\mathfrak g))$. Computing the central charge of this coset when $\mathfrak g$ is in the DC series gives $-\frac{22}{5}$ and indeed it turns out that \cite{ACKprin}
\[
\text{Com}\left(L_{-\frac{h^\vee}{6}}(\mathfrak g), L_{-\frac{h^\vee}{6}-1}(\mathfrak g) \otimes L_1(\mathfrak g)\right) \cong  M(5, 2).
\]
In other words, $L_{-\frac{h^\vee}{6}-1}(\mathfrak g) \otimes L_1(\mathfrak g)$ is a conformal extension of $L_{-\frac{h^\vee}{6}}(\mathfrak g) \otimes M(5, 2)$ and thus the Urod Theorem \cite{Arakawa:2020oqo} implies that $W_{-\frac{h^\vee}{6}-1}(\mathfrak{g}, f_\text{min}) \otimes L_1(\mathfrak g) \cong L_1(\mathfrak g)$ is a conformal extension of $M(5, 2) \otimes W_{-\frac{h^\vee}{6}}(\mathfrak{g}, f_\text{min})$.
For any $\mathfrak{g}$ in the DC series, we have a decomposition
\begin{equation}\label{eq decomposition L1(g)}
    \begin{split}
        L_1(\mathfrak g) &\cong A_\mathfrak g \otimes L_1(\mathfrak{sl_2}) \oplus B_\mathfrak g  \otimes L_1^{\mathfrak{sl}_2}(\omega) \\
        &\cong A_\mathfrak g \otimes \left(M(5, 2) \otimes M(5, 3) \oplus  M^{5, 2}_{3, 1} \otimes    M^{5, 3}_{3, 1}\right) \oplus  \\
        &\qquad  B_\mathfrak g\otimes \left(M^{5, 2}_{2, 1} \otimes    M^{5, 3}_{2, 1} \oplus  M^{5, 2}_{4, 1} \otimes    M^{5, 3}_{4, 1}  \right) \\
        &=  M(5, 2) \otimes \left( A_\mathfrak g \otimes M(5, 3) \oplus B_\mathfrak g\otimes M^{5, 3}_{4, 1}  \right) \oplus \\
        &\qquad M^{5, 2}_{3, 1} \otimes \left(  A_\mathfrak g \otimes M^{5, 3}_{3,1 } \oplus B_\mathfrak g \otimes M^{5, 3}_{2, 1}  \right)\ ,
    \end{split}
\end{equation}
where $A_\mathfrak g$ and $B_\mathfrak g$ for each $\mathfrak{g}$ are summarized in the Table \ref{tab:DC-decomposition}.
\begin{table}[t]
\centering
\begin{tabular}{c|c|c}
\hline
$\mathfrak g$ 
& $A_{\mathfrak g}$ 
& $B_{\mathfrak g}$ \\ 
\hline
$A_1$ 
& $\mathbf 1$ 
& $\mathbf 1$ \\[2pt]

$A_2$ 
& $V_{\sqrt{3}A_1}$ 
& $V_{\sqrt{3}A_1+\lambda/2}$ \\[2pt]

$G_2$ 
& $L_3(\mathfrak{sl}_2)$ 
& $L_3(3\omega)$ \\[2pt]

$D_4$ 
& $L_1(\mathfrak{sl}_2)^{\otimes 3}$ 
& $L_1(\omega)^{\otimes 3}$ \\[2pt]

$F_4$ 
& $L_1(\mathfrak{sp}_6)$ 
& $L_1^{\mathfrak{sp}_6}(\omega_3)$ \\[2pt]

$E_6$ 
& $L_1(\mathfrak{sl}_6)$ 
& $L_1^{\mathfrak{sl}_6}(\omega_3)$ \\[2pt]

$E_7$ 
& $L_1(\mathfrak{so}_{12})$ 
& $L_1^{\mathfrak{so}_{12}}(\omega_6)$ \\[2pt]

$E_8$ 
& $L_1(E_7)$ 
& $L_1^{\mathfrak{e}_7}(\omega_1)$ \\
\hline
\end{tabular}
\vspace{0.7em}
\caption{
Decomposition of $L_1(\mathfrak g)$ for $\mathfrak g$ in the DC series,
$L_1(\mathfrak g) \cong A_\mathfrak g \otimes L_1(\mathfrak{sl_2}) \oplus B_\mathfrak g  \otimes L_1(\omega)$. Here $V_L$ is the Lattice VOA associated to the lattice $L$.
}
\label{tab:DC-decomposition}
\end{table}
We then have
\be\label{commutant M52 L1g}
\text{Com}\left( M(5,2), L_1(\mathfrak g)\right) \cong  A_\mathfrak g \otimes M(5, 3) \oplus B_\mathfrak g \otimes M^{5, 3}_{4, 1}\ ,
\ee
which is naturally realized as a boundary vertex algebra of $\overline\CT^A_\text{min}\times T_\mathfrak{g}$. This class of VOAs is identified with the $C_2$-cofinite and $\mathbb{Z}_2$-rational $W$-algebra $W_{k}(\mathfrak{g}, f_\text{min})$ at level $k=-h^\vee/6$ \cite{kawasetsu2018algebras}.

The representation theory of the affine vertex superalgebras of $\mathfrak{osp}_{1|2n}$ behaves in many respects like the one of vertex algebras associated to simple Lie algebras \cite{Creutzig:2022riy}. In particular $L_m(\mathfrak{osp}_{1|2n})$ is also rational for any positive integer level $m$. The dual Coxeter number is $h^\vee = n+\frac{1}{2}$ and the superdimension is $\text{sdim}(\mathfrak{osp}_{1|2n})=n(2n-1)$. So that we see that 
\[
\text{sdim}(\mathfrak{osp}_{1|2n}) = \frac{2(h^\vee+1)(5h^\vee-6)}{h^\vee + 6}\ ,
\]
if and only if $n=1$. It turns out that also in this case $W_{k}(\mathfrak{g}, f_\text{min}) \cong \mathbb C$ is trivial \cite{Adamovic:2016jjs, Adamovic:2025ilx}.
The same proof as \cite{ACKprin} shows that 
\be
\text{Com}\left(L_{-\frac{1}{4}}(\mathfrak{osp}_{1|2}), L_{-\frac{5}{4}}(\mathfrak{osp}_{1|2}) \otimes L_1(\mathfrak{osp}_{1|2})\right) \cong  M(5, 2)
\ee
as well 
and hence by the Urod Theorem \cite{Arakawa:2020oqo}, which also holds for Lie superalgebras,
$L_1(\mathfrak{osp}_{1|2})$ is a conformal extension of $M(5, 2) \otimes W_{-\frac{1}{4}}(\mathfrak{osp}_{1|2}, f_{\text{min}})$.
The decomposition of $L_1(\mathfrak{osp}_{1|2})$ is \cite{Creutzig:2018zqd}
\begin{equation}
    \begin{split}
        L_1(\mathfrak{osp}_{1|2}) &\cong M(5, 3) \otimes L_1(\mathfrak{sl_2}) \oplus M^{5,3}_{4,1}  \otimes L_1^{\mathfrak{sl}_2}(\omega)
         \end{split}
\end{equation}
so that if we set
\[
 A_{\mathfrak{osp}_{1|2}} = M(5, 3), \qquad  B_{\mathfrak{osp}_{1|2}} = M^{5,3}_{4,1},
\]
then we get as before 
\begin{equation}
    \begin{split}
       L_1(\mathfrak{osp}_{1|2})  &\cong
        A_{\mathfrak{osp}_{1|2}} \otimes L_1(\mathfrak{sl_2}) \oplus B_{\mathfrak{osp}_{1|2}}  \otimes L_1^{\mathfrak{sl}_2}(\omega) \\
        &\cong A_{\mathfrak{osp}_{1|2}} \otimes \left(M(5, 2) \otimes M(5, 3) \oplus  M^{5, 2}_{3, 1} \otimes    M^{5, 3}_{3, 1}\right) \oplus  \\
        &\qquad  B_{\mathfrak{osp}_{1|2}} \otimes \left(M^{5, 2}_{2, 1} \otimes    M^{5, 3}_{2, 1} \oplus  M^{5, 2}_{4, 1} \otimes    M^{5, 3}_{4, 1}  \right) \\
        &=  M(5, 2) \otimes \left( A_{\mathfrak{osp}_{1|2}} \otimes M(5, 3) \oplus B_{\mathfrak{osp}_{1|2}} \otimes M^{5, 3}_{4, 1}  \right) \oplus \\
        &\qquad M^{5, 2}_{3, 1} \otimes \left(  A_{\mathfrak{osp}_{1|2}} \otimes M^{5, 3}_{3,1 } \oplus B_{\mathfrak{osp}_{1|2}} \otimes M^{5, 3}_{2, 1}  \right)\ .
    \end{split}
\end{equation}

The $W$-algebra of $\mathfrak{osp}_{1|2}$  is realized as the boundary algebra of $\overline{\CT}_\text{min}^A\times T_{\mathfrak{osp}_{1|2}}$, where the TFT $T_{\mathfrak{osp}_{1|2}}$ is identified with $\CT^B_{\text{min}}$ equipped with the standard Dirichlet boundary condition \cite{Ferrari:2023fez}. As discussed in \emph{loc. cit.}, the self-mirror property of $\CT_{\text{min}}$ implies that this is mirror dual to
\be
\CT^B_{\text{min}}~~\text{with}~~ (\CD,D) ~~\longleftrightarrow~~\overline{\CT}^A_{\text{min}}~~\text{with}~~ (\CN,N)~~\text{coupled to}~~(\rm{bc})^{\otimes 2} \ ,
\ee
where $(\rm{bc})^{\otimes 2}$ denotes two copies of the $\rm{bc}$ ghost system, i.e. (0,2) Fermi multiplet for a certain choice of $R$-charge. The bulk theory can then be described as $\overline{\CT}_\text{min}^A\times \overline{\CT}_\text{min}^A$ with the Neumann boundary condition $(\CN,N;\CN,N)$ coupled to $(\rm{bc})^{\otimes 2}$.

\subsection{Characters and supercharacters} For $\mathfrak{g} = D_4, E_6, E_7, E_8$, where $n_\theta^\vee$ is the flux through the Dynkin node attached to the affine node, we can write the Neumann half-index \eqref{half index L1g} as
\begin{equation}
\begin{split}
I^{\mathfrak{g}}_{\text{half}}(q;-1,1) =&~ \frac{1}{(q)^{\text{rk}(\mathfrak{g})-1}_\infty}\sum_{\substack{n\in \mathbb{Z}^{\text{rk}(\mathfrak{g})}\\n_{\theta^\vee}\geq 0}} \frac{1}{(q)_{n_{\theta^{\vee}}}} q^{\frac12 n^t C(\mathfrak{g}) n} (-q^{1/2})^{n_{\theta^\vee}}\\
=&~\frac{1}{(q)_\infty^{\text{rk}(\mathfrak g)-1}} \sum_{\substack{m\in \mathbb{Z}^{\text{rk}(\mathfrak{g})-1}\\n_{\theta^\vee}\geq 0}} \frac{1}{(q)_{n_{\theta^{\vee}}}} q^{n_{\theta^\vee}^2 + n_{\theta^\vee}\sum_{j\neq \theta^\vee} C(\mathfrak{g})_{\theta^\vee j }m_j+\frac12 m^t C(\mathfrak{g}_0' )m} (-q^{1/2})^{n_{\theta^\vee}}\ .
\end{split}
\end{equation}
This expression can be decomposed into a sum of even and odd $n_{\theta^\vee}$:
\begin{equation}
\begin{split}
I^{\mathfrak{g}}_{\text{half}}(q;-1,1)  =&~ \frac{1}{(q)_\infty^{\text{rk}(\mathfrak g)-1}} \sum_{\substack{m\in \mathbb{Z}^{\text{rk}(\mathfrak{g})-1}\\k\geq 0}} \frac{1}{(q)_{2k}} q^{4k^2+k + 2k \sum_{j\neq \theta^\vee} C(\mathfrak{g})_{\theta^\vee j }m_j +\frac12 m^t C(\mathfrak{g}_0' )m} \\
&-\frac{q^{1/2}}{(q)_\infty^{\text{rk}(\mathfrak g)-1}} \sum_{\substack{m\in \mathbb{Z}^{\text{rk}(\mathfrak{g})-1}\\k\geq 0}} \frac{1}{(q)_{2k+1}} q^{(2k+1)^2 +k+ ( 2k+1) \sum_{j\neq \theta^\vee} C(\mathfrak{g})_{\theta^\vee j }m_j +\frac12 m^t C(\mathfrak{g}_0' )m}\ .
\end{split}
\end{equation}
Let $u=-\sum_{j\neq \theta^\vee} C_{\theta^\vee j}\omega_j'$, where $\omega_j'$ is the $j$-th fundamental weight of $\mathfrak g_0'$. Shifting $m\rightarrow m + 2ku$, we have
\begin{equation}
\begin{split}
I^{\mathfrak{g}}_{\text{half}}(q;-1,1)  =&~ \left(\sum_{k\geq 0} \frac{1}{(q)_{2k}} q^{k^2+k} \right) \frac{1}{(q)_\infty^{\text{rk}(\mathfrak g_0')}} \sum_{m\in \mathbb{Z}^{\text{rk}(\mathfrak{g}_0')}} q^{ \frac12 m^t C(\mathfrak{g}_0' )m}\\
&~-q^{1/2}   \left(\sum_{k\geq 0} \frac{1}{(q)_{2k+1}} q^{(k+1)^2} \right) \frac{1}{(q)_\infty^{\text{rk}(\mathfrak g_0')}} \sum_{m\in \mathbb{Z}^{\text{rk}(\mathfrak{g}_0')}} q^{ \frac12 m^t C(\mathfrak{g}_0' )m + \sum_{j\neq \theta^\vee} C(\mathfrak{g})_{\theta^\vee j} m_j}\ .
\end{split}
\end{equation}
This is precisely the supercharacter of \eqref{commutant M52 L1g}, up to a modular anomaly prefactor.

The cases $\mathfrak{g}=A_2, G_2, F_4$ require separate treatment. For $\mathfrak{g}=A_2$, 
\begin{equation}
\begin{split}
I^{A_2}_{\text{half}}(q;-1,1) =&~ \frac{1}{(q)_\infty} \sum_{\substack{(n_1,n_2)\in \mathbb{Z}^2\\n_1+n_2\geq 0}} \frac{1}{(q)_{n_1+n_2}} q^{n_1^2+n_2^2 - n_1n_2} (-q^{1/2})^{n_1+n_2} \\
=&~  \frac{1}{(q)_\infty} \sum_{n\geq 0, n_2 \in\mathbb{Z}} \frac{1}{(q)_n} q^{n^2 - 3n n_2 + 3n_2^2} (-q^{1/2})^{n} \\
=&~ \frac{1}{(q)_\infty} \sum_{k\geq 0, n_2 \in\mathbb{Z}} \frac{1}{(q)_{2k}} q^{4k^2 +k- 6k n_2 + 3n_2^2} \\
&~ -  \frac{q^{1/2}}{(q)_\infty} \sum_{k\geq 0, n_2 \in\mathbb{Z}} \frac{1}{(q)_{2k+1}} q^{(2k+1)^2 +k- 3(2k+1) n_2 + 3n_2^2} \ .
\end{split}
\end{equation}
Shifting $n_2\rightarrow n_2+k$,
\begin{equation}
\begin{split}
I^{A_2}_{\text{half}}(q;-1,1) =&~ \left( \sum_{k\geq 0} \frac{1}{(q)_{2k}} q^{k^2 +k}\right) \frac{1}{(q)_\infty} \sum_{n_2\in \mathbb{Z}} q^{3n_2^2 } \\
&~-q^{-1/4}\left( \sum_{k\geq 0} \frac{1}{(q)_{2k+1}} q^{(k+1)^2}\right) \frac{1}{(q)_\infty} \sum_{n_2\in \mathbb{Z}+\frac12} q^{3n_2^2}\ ,
\end{split}
\end{equation}
which coincides with the supercharacter of \eqref{commutant M52 L1g} for $\mathfrak g= A_2$, up to a modular anomaly prefactor. In this case, $A_\mathfrak{g}$ is the lattice VOA $V_{\sqrt{3}A_1}$ associated with the lattice $\sqrt3 A_1 = \lambda\mathbb{Z}$ with $(\lambda,\lambda)=6$, and $B_\mathfrak{g}$ is the module $V_{\sqrt{3}A_1 + \lambda/2}$.

Finally, for $\mathfrak{g}= G_2,F_4$, we expand the first line of \eqref{half index L1g} as a $q$-series and check that the first few coefficients match with the supercharacters of \eqref{commutant M52 L1g}.

\subsection{\texorpdfstring{$HT^B$}{HTB}-twist and intermediate vertex subalgebras} The Neumann half-index for the $HT^B$-twisted theory is 
\begin{equation}\label{HTB N half index}
I^{\mathfrak{g}}_{\text{half}}(q;1,1) = (q)_\infty \oint \frac{dz}{2\pi i z} \frac{1}{(z;q)_\infty}I^{\mathfrak{g}}_{2d}(q;z,1,1)\ .
\end{equation}
For simply-laced $\mathfrak{g}$, this is evaluated as 
\begin{equation}
I^{\mathfrak{g}}_{\text{half}}(q;1,1) = \frac{1}{(q)^{\text{rk}(\mathfrak{g})-1}_\infty}\sum_{\substack{n\in \mathbb{Z}^{\text{rk}(\mathfrak{g})}\\ n_{\theta^\vee}\geq 0}} \frac{1}{(q)_{n_{\theta^{\vee}}}} q^{\frac12 n^t C(\mathfrak{g}) n}\ ,
\end{equation}
which coincides with the conjectural expressions for the characters of the intermediate vertex subalgebras computed in \cite{Lee:2024fxa}, as summarized in Table \ref{IVAs}. For non-simply-laced $\mathfrak{g}$, we compute the first few coefficients of \eqref{HTB N half index} and verify that they agree with the corresponding character expansions of the intermediate vertex algebras (IVAs) in \emph{loc. cit.}\footnote{The remaining exotic case, denoted $D_{6+1/2+1/2}$ in \cite{Lee:2024fxa} can be  realized in a similar manner by coupling two copies of $\overline{\CT}_{\text{min}}^A$ with $L_1(\mathfrak{e}_8)$.}

\begin{table}[h]
\centering
\begin{tabular}{c|cccccccc}
$\mathfrak g$ 
& $A_1$ & $A_2$ & $G_2$ &$D_4$ & $F_4$ & $E_6$ & $E_7$ & $E_8$ \\
\hline
IVA & $IM$ & $UA_{1+1/2}$ & $AG_{1+1/2}$ & $AD_{3+1/2}$  & $C_{3+1/2}$ &  $A_{5+1/2}$& $D_{6+1/2}$& $E_{7+1/2}$
\end{tabular}
\vspace{0.7em}
\caption{Intermediate vertex subalgebras}
\label{IVAs}
\end{table}

The boundary operators survive the $Q$-cohomology are
\be
J^\alpha_{\mathfrak{g}_0'}\ , \quad :\phi^a J_{(-a)}: \text{ for } a=1,2\ ,
\ee
where $J_{(a)}$ collectively denote the $L_1(\mathfrak{g})$-currents with $h_\theta$-charge $a$. All of these operators have twisted spin 1, forming an intermediate Lie algebra $\mathfrak{g}_0'\oplus \mathfrak{g}_1\oplus \mathfrak{g}_2$ inside $\mathfrak{g}$. The OPEs among them are directly inherited from those of $L_1(\mathfrak{g}_0')$. The operator $\phi^2 J_{(-2)}$ has regular OPEs with all generators, and can be identified with the boundary image of the bulk monopole operator $\phi^2 V_{-1}$.

\section{Mirror descriptions} \label{sec: mirror}

\subsection{Dual descriptions}
When $\mathfrak{g}$ is simply-laced, we introduce another family of $\CN=2$ gauge theories, denoted by
$\mathbb{T}_{\mathfrak g}$, and conjecture that they flow in the infrared to ${\CT}_{\text{min}}\times T_{\mathfrak g}$, the $\CN=4$ minimal SCFT coupled to the TFT $T_{\mathfrak g}$.
Since $\CT_{\text{min}}$ is dual to its orientation reversal under 3d $\CN=4$ mirror symmetry \cite{Creutzig:2024ljv}, this may be viewed as a mirror-dual description of $\overline{\CT}_\text{min}\times T_\mathfrak g$.

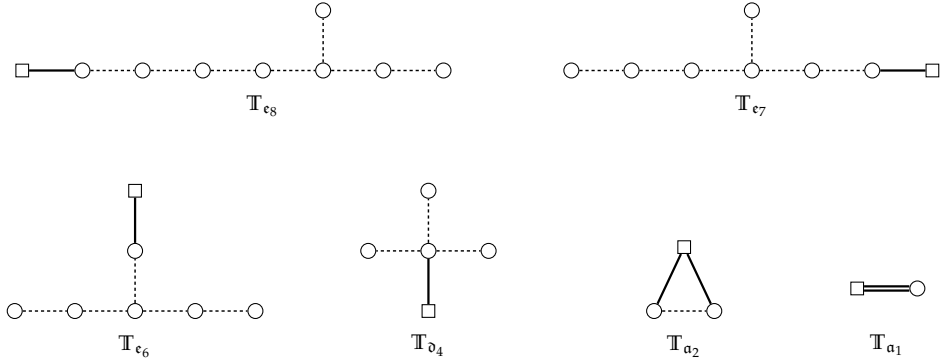
\begin{figure}[t]
  \centering
\begin{tikzpicture}[
  baseline={(current bounding box.center)},
  node distance=0.80cm,
  circ/.style={circle, draw, inner sep=2.0pt},
  sq/.style={draw, rectangle, inner sep=2.4pt},
  dynedge/.style={line width=0.55pt, dash pattern=on 1.2pt off 1.1pt},
  thicklink/.style={line width=0.85pt}
]

\def\COL{4.8cm}     
\def\ROW{3.2cm}     
\def\D4RAISE{0.8cm}

\coordinate (B1) at (0,0);
\coordinate (B2) at (\COL,0);
\coordinate (B3) at (2*\COL,0);
\coordinate (B4) at (3*\COL,0);

\coordinate (T1) at (0.5*\COL,\ROW);     
\coordinate (T2) at (2.5*\COL,\ROW);     

\begin{scope}[shift={(T1)}, xshift=-1.5cm]
  \node[circ] (e8_1) {};
  \node[circ] (e8_2) [right of=e8_1] {};
  \node[circ] (e8_3) [right of=e8_2] {};
  \node[circ] (e8_4) [right of=e8_3] {};
  \node[circ] (e8_5) [right of=e8_4] {};
  \node[circ] (e8_6) [right of=e8_5] {};
  \node[circ] (e8_7) [right of=e8_6] {};
  \node[circ] (e8_v) [above of=e8_5] {};
  \node[sq]   (e8_sq) [left of=e8_1] {};

  \draw[dynedge] (e8_1)--(e8_2)--(e8_3)--(e8_4)--(e8_5)--(e8_6)--(e8_7);
  \draw[dynedge] (e8_5)--(e8_v);
  \draw[thicklink] (e8_sq)--(e8_1);

  \node[below=6pt of e8_4, font=\scriptsize, draw=none, shape=rectangle, inner sep=0pt]
    {$\mathbb{T}_{\mathfrak{e}_8}$};
\end{scope}

\begin{scope}[shift={(T1)}, xshift=5cm]
  \node[circ] (e7_1) {};
  \node[circ] (e7_2) [right of=e7_1] {};
  \node[circ] (e7_3) [right of=e7_2] {};
  \node[circ] (e7_4) [right of=e7_3] {};
  \node[circ] (e7_5) [right of=e7_4] {};
  \node[circ] (e7_6) [right of=e7_5] {};
  \node[circ] (e7_7) [above of=e7_4] {};
  \node[sq]   (e7_sq) [right of=e7_6] {};

  \draw[dynedge] (e7_1)--(e7_2)--(e7_3)--(e7_4)--(e7_5)--(e7_6);
  \draw[dynedge] (e7_4)--(e7_7);
  \draw[thicklink] (e7_6)--(e7_sq);

  \node[below=6pt of e7_4, font=\scriptsize, draw=none, shape=rectangle, inner sep=0pt]
    {$\mathbb{T}_{\mathfrak{e}_7}$};
\end{scope}

\begin{scope}[shift={(B1)}]
  \node[circ] (e6_1) {};
  \node[circ] (e6_2) [right of=e6_1] {};
  \node[circ] (e6_3) [right of=e6_2] {};
  \node[circ] (e6_4) [right of=e6_3] {};
  \node[circ] (e6_5) [right of=e6_4] {};
  \node[circ] (e6_t) [above of=e6_3] {};
  \node[sq]   (e6_sq) [above of=e6_t] {};

  \draw[dynedge] (e6_1)--(e6_2)--(e6_3)--(e6_4)--(e6_5);
  \draw[dynedge] (e6_3)--(e6_t);
  \draw[thicklink] (e6_t)--(e6_sq);

  \node[below=6pt of e6_3, font=\scriptsize, draw=none, shape=rectangle, inner sep=0pt]
    {$\mathbb{T}_{\mathfrak{e}_6}$};
\end{scope}

\begin{scope}[shift={(B1)}, xshift=5.5cm]
\begin{scope}[yshift=\D4RAISE]
  \node[circ] (d4_c) {};
  \node[circ] (d4_l) [left of=d4_c] {};
  \node[circ] (d4_r) [right of=d4_c] {};
  \node[circ] (d4_u) [above of=d4_c] {};
  \node[sq]   (d4_sq) [below of=d4_c] {};

  \draw[dynedge] (d4_c)--(d4_l);
  \draw[dynedge] (d4_c)--(d4_r);
  \draw[dynedge] (d4_c)--(d4_u);
  \draw[thicklink] (d4_c)--(d4_sq);

  \node[below=6pt of d4_sq, font=\scriptsize, draw=none, shape=rectangle, inner sep=0pt]
    {$\mathbb{T}_{\mathfrak{d}_4}$};
\end{scope}
\end{scope}

\begin{scope}[shift={(B1)}, xshift=8.5cm]
  \node[circ] (a2_1) {};
  \node[circ] (a2_2) [right of=a2_1] {};
  \node[sq]   (a2_sq) at ($(a2_1)!0.5!(a2_2)+(0,0.85cm)$) {};

  \draw[dynedge] (a2_1)--(a2_2);
  \draw[thicklink] (a2_sq)--(a2_1);
  \draw[thicklink] (a2_sq)--(a2_2);

  \node[below=6pt of a2_1, xshift=0.40cm, font=\scriptsize, draw=none, shape=rectangle, inner sep=0pt]
    {$\mathbb{T}_{\mathfrak{a}_2}$};
\end{scope}

\begin{scope}[shift={(B1)}, xshift=12cm, yshift= 0.3cm]
  \node[circ] (a1_c) {};
  \node[sq]   (a1_sq) [left of=a1_c] {};

  \draw[thicklink] ([yshift=0.8pt]  a1_sq.east) -- ([yshift=0.8pt]  a1_c.west);
  \draw[thicklink] ([yshift=-0.8pt] a1_sq.east) -- ([yshift=-0.8pt] a1_c.west);

\node[
  below=17pt,
  font=\scriptsize,
  draw=none,
  shape=rectangle,
  inner sep=0pt
] at ($(a1_sq)!0.5!(a1_c)$)
  {$\mathbb{T}_{\mathfrak{a}_1}$};
\end{scope}
\end{tikzpicture}

  \caption{Quiver diagrams for the $\CN=2$ gauge theory description for $\mathbb{T}_\mathfrak{g}$. Circular nodes denote $U(1)$ gauge groups, interacting through the CS couplings with level matrix given by $K=C(\mathfrak{g})$, represented by dotted lines. Solid lines together with square nodes denote chiral multiplets of unit charge. In case of $\mathbb{T}_{\mathfrak{a}_1}$, a double solid line indicates a chiral multiplet of charge 2. These complete the quivers to the $\mathfrak{g}$-affine Dynkin diagrams.}
  \label{affine quiver}
\end{figure}

The theory $\mathbb{T}_{\mathfrak g}$ admits a $\CN=2$ Lagrangian description as a
  $G$ Chern-Simons theory at level one, coupled to a single chiral multiplet charged under $U(1)\subset G$ with unit charge, specified by $\rho\in\mathfrak{h}^*$, as listed in Table \ref{weight rho}.
\begin{table}[t]
\centering
\begin{tabular}{c|cccccc}
$\mathfrak g$ 
& $A_1$ & $A_2$ & $D_4$ & $E_6$ & $E_7$ & $E_8$ \\
\hline
$\rho$ 
& $2\omega$ & $\omega_1 + \omega_2$  & $\omega_2$  &  $\omega_6$ & $\omega_1$ & $\omega_1$  \\
\end{tabular}
\vspace{0.7em}
\caption{
Charges of the chiral multiplet in $\mathbb{T}_{\mathfrak g}$.
}
\label{weight rho}
\end{table}
This theory is alternatively realized as the $\CN=2$ $U(1)_K^{\text{rk}(\mathfrak g)}$ Chern-Simons theory with the level matrix $K= C(\mathfrak{g})$, coupled to a chiral multiplet with charge $\rho\in \mathbb{Z}^{\text{rk}(\mathfrak g)}$.\footnote{Here the CS level $K$ is the bare CS levels in "$U(1)_{\mathfrak{-1/2}}$ quantization", as defined in \cite{Closset:2018ghr}. The UV effective level is $k=K-\frac12 \rho^T\rho$.}  It can be represented as an abelian quiver Chern-Simons theory, in which the chiral multiplet completes the quiver to the $\mathfrak{g}$-affine Dynkin diagram, as in Figure \ref{affine quiver}.

Notice that the quiver diagram for $\mathbb{T}_{\mathfrak{a}_1}$ describes an $\CN=2$ $U(1)_2$ CS theory coupled to a chiral multiplet of charge 2, a theory first considered in \cite{Gaiotto:2018yjh}. In Appendix B of \emph{loc. cit.}, the authors establish the duality between $\mathbb{T}_{\mathfrak{a}_1}$ and $\CT_{\text{min}}\times U(1)_2$\footnote{In \cite{Gaiotto:2018yjh}, they adopt the UV effective CS level, where $\mathbb{T}_{\mathfrak{a}_1}$ and is described as $U(1)_0$ coupled to a charge-2 chiral multiplet.}, which is consistent with our proposal. The quiver theory and its boundary algebra for $\mathbb{T}_{\mathfrak{e}_8}$ appear in \cite{Kim:2024dxu} as the vertex algebra associated with the fourth power of the BPS monodromy operator of the $(A_1,A_2)$-Argyres Douglas theory.

\subsection{Supersymmetry enhancement}

The gauge theory $\mathbb{T}_{\mathfrak{g}}$ has a $U(1)^{\text{rk}(\mathfrak{g})}$ topological symmetry, but all of them decouple in the infrared except for a single $U(1)$ global symmetry. We conjecture that this symmetry is naturally identified with the axial symmetry $S$ in the infrared, which can be represented as 
\be
S = -\frac12 \sum_{i,j}a_i^\vee C(\mathfrak{g})_{ij} M_j\ ,
\ee
where $M_j$ is the $U(1)$ topological symmetry associated with $j$-th gauge node. With this identification, we find that the gauge theory $\mathbb{T}_\mathfrak g$ possesses two gauge invariant $1/4$-BPS dressed monopole operators 
\be
\phi V_{-\theta}\ ,\qquad \phi V_{\theta}\ ,
\ee
with superconformal $R$-charge 1, spin $1$ and axial charge $S=\pm 1$, where $\theta$ is the highest-root vector in the simple root basis. This provides strong evidence that these operators belong to the extra supercurrent multiplet associated with $\CN=4$ supersymmetry enhancement in the infrared SCFT sector. Performing $F$-maximization and computing the superconformal indices and the partition functions on Seifert manifolds for $\mathbb{T}_\mathfrak g$, we find that they match those of $\CT_{\text{min}}\times T_\mathfrak g$, consistent with the proposal.

\subsection{Dual boundary conditions}
In the previous sections, we analyzed the $(\CN, N; \CD)$ boundary condition for $\overline{\CT}_{\text{min}}\times T_{\mathfrak g}$, i.e. a Neumann boundary condition for $\overline{\CT}_{\text{min}}$ together with Dirichlet boundary conditions for $T_{\mathfrak g}$, coupled as described in Section \ref{subsec: boundary conditions}. We propose that the dual boundary condition for $\mathbb{T}_{\mathfrak g}$ is a deformed Dirichlet (with the non-zero boundary value for the chiral multiplet), so that 3d mirror symmetry exchanges
\be\label{mirror boundary}
\overline{\CT}_{\text{min}}\times T_{\mathfrak g} ~\text{ with }~ (\CN, N; \CD) ~\longleftrightarrow ~\mathbb{T}_{\mathfrak g} ~\text{ with }~ (\CD,D_c)\ .
\ee

 This claim is supported by a calculation of the half-index. The Dirichlet half-index of $\mathbb{T}_{\mathfrak g}$ is
\begin{equation}
I^{{\mathbb{T}_\mathfrak{g}}}_{\text{half}}(q;\nu,\eta,s) 
= \frac{1}{(q)^{\text{rk}(\mathfrak{g})-1}_\infty}\sum_{\substack{n\in \mathbb{Z}^{\text{rk}(\mathfrak{g})}\\n_{\theta^\vee}\geq 0}}  \frac{q^{\frac12 n^t C(\mathfrak{g}) n}}{(q)_{n_{\theta^\vee}}}\left[(-q^{1/2})^{\nu+1}\eta\right]^{n_{\theta^\vee}/2} \prod_{i,j=1}^{\text{rk}(\mathfrak{g})-1}s_{i}^{-C(\mathfrak{g}_0')_{ij}n_j} \ .
\end{equation}
Comparing this expressions with \eqref{half index L1g} (with a straightforward refinement of the Jacobi variables for $\mathfrak g_0'$), we find
\be
I^{\mathfrak{g}}_{\text{half}}(q;\nu,\eta,s) = I^{{\mathbb{T}_\mathfrak{g}}}_{\text{half}}(q;-\nu,\eta^{-1},s) \ ,
\ee
which provides strong evidence for the proposal \eqref{mirror boundary}.

It is worth noting that the form of the $1/4$-BPS monopole operators above is further indication that this deformed Dirichlet boundary condition is compatible with the $B$-twist, but not the $A$-twist.
Namely, as indicated by the above index, a Dirichlet boundary condition with $\phi \neq 0$ will lift all monopole operators with $n_{\theta^\vee} < 0$ as described in Section 5.1 of \cite{Ferrari:2023fez} (see Footnote 9).
In the $HT$ twist, this implies the restriction of $\phi V_{-\theta}$ vanishes on the boundary, whereas $\phi V_{\theta}$ is nonzero.
In other words, this boundary condition is deformable to the $B$-twist \cite{Brunner:2021tfl, Ferrari:2023fez}.
In light of the mirror symmetry proposed in Eq. \eqref{mirror boundary}, we conclude that the boundary conditions studied in the previous sections are deformable to the $A$-twist, but not the $B$-twist.

\bibliography{ADLR}
\bibliographystyle{amsalpha}

\end{document}